\documentclass{ws-ijmpa}

\begin{document}
\def \ee {{\varepsilon}}
\def \ez {{\varepsilon_0}}
\def \ef {{\varepsilon^{(1)}}}
\def \es {{\varepsilon^{(2)}}}
\def \ezf {{\varepsilon_0^{(1)}}}
\def \ezs {{\varepsilon_0^{(2)}}}
\def \rf {{r_0^{(1)}}}
\def \rs {{r_0^{(2)}}}
\def \rz {{r_0}}
\markboth{B.~Geyer, G.~L.~Klimchitskaya and V.~M.~Mostepanenko}
{Thermal Casimir force between dielectrics}

%
\catchline{}{}{}{}{}
%

\title{RECENT RESULTS ON THERMAL CASIMIR FORCE BETWEEN DIELECTRICS AND 
RELATED PROBLEMS
}
\thispagestyle{empty}

\author{B.~GEYER, G.~L.~KLIMCHITSKAYA\footnote{On leave from
North-West Technical University, St.Petersburg, Russia.}{\ } and 
V.~M.~MOSTEPANENKO\footnote{On leave from Noncommercial
Partnership ``Scientific Instruments'',
Moscow,  Russia.} 
}

\address{Center of Theoretical Studies and Institute for Theoretical Physics, 
Leipzig University,\\
Augustusplatz 10/11, 04109, Leipzig, Germany}

\maketitle

\pub{Received 26 May 2006}{}

\begin{abstract}
We review recent results obtained in the physics of the 
thermal Casimir force acting between two dielectrics, 
dielectric and metal, and between metal and semiconductor. 
The detailed derivation for the low-temperature behavior 
of the Casimir free energy, pressure and entropy in the 
configuration of two real dielectric plates is presented.
For dielectrics with finite static dielectric permittivity 
it is shown that the Nernst heat theorem is satisfied. 
Hence, the Lifshitz theory of the van der Waals and 
Casimir forces is demonstrated to be consistent with 
thermodynamics. The nonzero dc conductivity of dielectric 
plates is proved to lead to a violation of the Nernst 
heat theorem and, thus, is not related to the phenomenon 
of dispersion forces. The low-temperature asymptotics of 
the Casimir free energy, pressure and entropy are derived 
also in the configuration of one metal and one dielectric 
plate. The results are shown to be consistent with
thermodynamics if the dielectric plate possesses a finite 
static dielectric permittivity. If the dc conductivity
of a dielectric plate is taken into account this results 
in the violation of the Nernst heat theorem. We discuss 
both the experimental and theoretical  results related 
to the Casimir interaction between metal and semiconductor
with different charge carrier density. Discussions in the 
literature on the possible influence of spatial dispersion 
on the thermal Casimir force are analyzed. In conclusion, 
the conventional Lifshitz theory taking into account only 
the frequency dispersion remains the reliable foundation 
for the interpretation of all present experiments.

\end{abstract}

\section{INTRODUCTION}

The Casimir effect\cite{1} is the force and also the specific polarization of
the vacuum arising in restricted quantization volumes and originating
from the zero-point oscillations of quantized fields. This force acts
between two closely spaced macrobodies, between an atom or a molecule
and macrobody or between two atoms or molecules. During more than fifty
years, passed after the discovery of the Casimir effect, it has
attracted much theoretical attention because of numerous applications
in quantum field theory, atomic physics, condensed matter physics,
gravitation and cosmology, mathematical physics, and in nanotechnology
(see monographs \refcite{2}--\refcite{4} and 
reviews \refcite{5}--\refcite{7}). In multidimensional
Kaluza-Klein supergravity the Casimir effect was used\cite{8} as a
mechanism for spontaneous compactification of extra spatial dimensions
and for constraining the Yukawa-type corrections to Newtonian
gravity.\cite{9,10,10a,10b,10c} 
In quantum chromodynamics the Casimir energy plays
an important role in the bag model of hadrons.\cite{3} In cavity quantum
electrodynamics the Casimir interaction between an isolated atom and
a cavity wall leads to the level shifts of atomic electrons depending
on the position  of the atom near the wall.\cite{2} Both the van der Waals
and Casimir forces are used\cite{11,12} for the theoretical interpretation
of recent experiments on quantum reflection and Bose-Einstein
condensation of ultracold atoms on or near the cavity wall of different
nature. In condensed matter physics the Casimir effect turned out to
be important for interaction of thin films, in wetting processes, and
in the theory of colloids and lattice defects.\cite{13} The Casimir force
was used to actuate nanoelectromechanical devices\cite{14} and to study
the absorption of hydrogen atoms by carbon nanotubes.\cite{15}
Theoretical work on the calculation of the Casimir energies and
forces stimulated important achievements in mathematical physics and
in the theory of renormalizations connected with the method of
generalized zeta function and heat kernel expansion.\cite{6,15a} All this
made the Casimir effect the subject of general interdisciplinary
interest and attracted permanently much attention in the scientific
literature.

The last ten years were marked by the intensive experimental investigation
of the Casimir force between metallic test bodies (see
Refs.~\refcite{16}--\refcite{28a}). During
this time the agreement between experiment and theory on the level of
1-2\% of the measured force was achieved. This has become  possible due 
to the use of modern laboratory techniques, in particular, of atomic force
microscopes and micromechanical torsional oscillators. Metallic test
bodies provide advantage in comparison with dielectrics because their
surfaces avoid charging. In Refs.~\refcite{Widom1,Widom2}, where the
importance of the Casimir effect for nanotechnology was pioneered, 
it was demonstrated that at 
separations below 100\,nm the Casimir force becomes larger than 
the typical electrostatic 
forces acting between the elements of microelectromechanical systems.
Bearing in mind that the miniaturization  is the main tendency in modern
technology, it becomes clear that the creation of new generation of
nanotechnological devices with further decreased elements and separations
between them
would become impossible without careful account and calculation of
the Casimir force.

Successful developments of nanotechnologies based on the Casimir effect
calls for more sophisticated calculation methods of the Casimir forces.
Most of theoretical output produced during the first decades after
Casimir's discovery did not take into account experimental conditions
and real material properties of the boundary bodies, such as surface
roughness, finite conductivity and nonzero temperature. The basic theory
giving the unified description of both the van der Waals and Casimir
forces was elaborated by Lifshitz\cite{29,29a,30} shortly after the publication
of Casimir's paper. It describes the boundary bodies in terms of the 
frequency dependent dielectric permittivity $\varepsilon(\omega)$ at
nonzero temperature $T$. In the applications of the Lifshitz theory
to dielectrics it was supposed that the static dielectric permittivity
(i.e., the dielectric permittivity at zero frequency) is finite. The
case of ideal metals was obtained from the Lifshitz theory by using
the so-called Schwinger's prescription,\cite{31} i.e., that the limit
$|\varepsilon(\omega)|\to\infty$ should be taken first and the static
limit $\omega\to 0$ second. For ideal metals the same result, as
follows from the Lifshitz theory combined with the Schwinger's
prescription, was obtained independently in the framework  of thermal
quantum field theory in Matsubara formulation.\cite{32} However, the cases
of real dielectrics and metals (which possess some nonzero dc
conductivity at $T>0$ and finite dielectric permittivity
at nonzero frequencies, respectively)
remained practically unexplored for a long time. The case of
semiconductor boundary bodies was also unexplored despite of the
crucial role of semiconductor materials in nanotechnology.

Starting in 2000, several theoretical groups in different countries
attempted to describe the Casimir interaction between real metals at
nonzero temperature in the framework of Lifshitz theory. They have
used different models of the metal conductivity and arrived to
controversial conclusions. In Ref.~\refcite{33}, using the dielectric
function of the Drude model, quite different results than for ideal
metals were obtained. According to Ref.~\refcite{33}, at short separations
(low temperatures) the thermal correction to the Casimir force acting
between real metals is several hundred times larger than between
ideal ones. In addition, at large separations of a few micrometers
(high temperatures) a two times smaller magnitude of the thermal
Casimir force was found than between ideal metals (the latter
is known as ``the classical limit''\cite{34,34a}). 
In Refs.~\refcite{35},\ \refcite{36} the
dielectric permittivity of the plasma model was used to describe
real metals and quite different results were obtained. At short
separations the thermal correction appeared to be small in
qualitative agreement with the case of ideal metals. At large
separations for real metals the familiar classical limit was
reproduced. Later the approach of Ref.~\refcite{33} was supported in
Refs.~\refcite{37},\ \refcite{38}. 
The plasma model approach can be used at such 
separations $a$ that the characteristic frequency $c/(2a)$ belongs to
the region of infrared optics. Later a more general framework, 
namely the impedance
approach was suggested\cite{39,40} which is applicable at any separation
larger than the plasma wave length.
It was supported in Refs.~\refcite{7},\ \refcite{41},\ \refcite{42}. 
In the region of the infrared
optics, the impedance approach leads to practically the same
results as the plasma model approach. As was shown in 
Refs.~\refcite{43},\ \refcite{44},
the Drude model approach leads to the violation of the Nernst heat
theorem when applied to perfect metal crystal lattices with no
impurities. This approach was also excluded by experiment at 99\%
confidence in the separation region from 300 to 500\,nm
and at 95\% confidence in the wider separation region from
170 to 700\,nm.\cite{26,28,28a} 
On the contrary, the plasma model and impedance
approaches were shown to be in agreement with thermodynamics and consistent
with experiment. The polemic between different theoretical approaches to
the description of the thermal Casimir force in the case of real
metals can be found in Refs.~\refcite{38},\ \refcite{45}--\refcite{45b}.

These findings on the application of the Lifshitz theory to real
metals have inspired a renewed interest  in the Casimir force between
dielectrics. As was mentioned above, at nonzero temperature
dielectrics possess an although small but not equal to zero dc conductivity.
In Ref.~\refcite{46},\ \refcite{46a} 
the van der Waals force arising from the dc conductivity
of a dielectric plate was shown to lead to large effect in noncontact 
atomic friction,\cite{47} a phenomenon having so far no satisfactory
theoretical explanation. This brings up the question: Is it
necessary or possible to take into account the dc conductivity of
dielectrics in the Lifshitz theory? Recall that in the case of a positive
answer the static dielectric permittivity of a dielectric material
would be infinitely large. It is amply clear that the resolution of the
above issue should be in accordance with the fundamentals of
thermodynamics. For this reason, it is desirable to investigate the
low-temperature behavior of the Casimir free energy and entropy for
two dielectric plates both with neglected and included effects of 
the dc conductivity.

A major breakthrough in the investigation of this problem was achieved in 
the year 2005. In Ref.~\refcite{48} a new variant of perturbation theory
was developed in a small parameter proportional to the product of
the separation distance between the plates and the temperature.
As a result, the behavior of the Casimir free energy, entropy and
pressure at low temperatures was found analytically. If the static
dielectric permittivity is finite, the thermal correction was
demonstrated to be in accordance with thermodynamics. This solves
positively the fundamental problem about the agreement between the
Lifshitz theory and thermodynamics for the case of two dielectric
plates. In Ref.~\refcite{48} it was shown that, on the contrary, the formal
inclusion of a small conductivity of dielectric plates at low
frequencies into the model of their dielectric response leads to a
 violation of the Nernst heat theorem. This result gives an important
guidance on how to extrapolate the tabulated optical data for the
complex refractive index to low frequencies in numerous applications
of the van der Waals and Casimir forces. All these problems and related
ones arising for semiconductor materials are discussed in this
review.

In Sec.~2 we derive the analytical behavior of the Casimir free energy,
entropy and pressure in the configuration of two parallel dielectric
plates at both low and high temperature. 
It is demonstrated that if the static dielectric permittivity
is finite the Lifshitz theory is in agreement with thermodynamics.
Sec.~3 contains the derivation of the low-temperature behavior for
the Casimir free energy and entropy between 
two dielectric plates with included dc conductivity. In this case the
Lifshitz theory is found to be in contradiction with the Nernst heat
theorem. The conclusion is made that the conductivity properties of
a dielectric material at a constant current are unrelated to the
van der Waals and Casimir forces and must not be included into the
model of dielectric response. In Sec.~4 we consider the thermal Casimir
force between metal and dielectric. This problem was first investigated
in Ref.~\refcite{49}. It was found that for dielectric plate with finite 
static dielectric permittivity the Nernst heat theorem is satisfied but 
the Casimir entropy may take negative values. Here we not only reproduce
an analytical proof of the Nernst heat theorem 
but also find the next perturbation orders in the
expansion of the Casimir free energy and entropy in powers of a small
parameter. The results obtained in high-temperature limit are also
provided. Sec.~5 is devoted to the Casimir interaction between metal
and dielectric plates with included dc conductivity of the dielectric
material. We demonstrate that in this case the Nernst heat theorem is
violated. Sec.~6 contains the discussion of semiconductors which present
a wide variety of electric properties varying from metallic to dielectric.
We consider the Casimir interaction between metal and semiconductor test
bodies and formulate the criterion when it is appropriate to include
the dc conductivity of a semiconductor into the model of dielectric
response. In doing so, the results of recent experiments\cite{50,51} on
the measurement of the Casimir force between metal and semiconductor 
test bodies are
taken into account. To this point the assumption has been made that
metal, dielectric and semiconductor materials of the Casimir plates
possess only temporal dispersion, i.e., can be described by the
dielectric permittivity depending only on frequency. In Sec.~7 we discuss
recent controversial results by different authors (see, e.g.,
Refs.~\refcite{52} -- \refcite{56}) 
attempting  to take into account also spatial dispersion.
As is shown in this section, the way of inclusion of spatial
dispersion into the Lifshitz theory, used in 
Refs.~\refcite{52} -- \refcite{56}, is
unjustified. We argue that the account of spatial dispersion cannot
influence theoretical results obtained with the help of usual,
spatially local, Lifshitz theory within presently used ranges of 
experimental separations. Sec.~8 contains our conclusions and discussion. 

\section{NEW ANALYTICAL RESULTS FOR THE THERMAL CASIMIR FORCE
BETWEEN DIELECTRICS}

We consider two thick dielectric plates (semispaces) described by the
frequency-dependent dielectric permittivity $\varepsilon(\omega)$ and
restricted by the parallel planes $z=\pm a/2$ with a separation $a$
between them, in thermal equilibrium at temperature $T$.
The Lifshitz formula for the free energy of the van der Waals and
Casimir interaction between the plates is given by\cite{2,3,4,5,6,29,29a,30}
\begin{eqnarray}
&&{\cal F}(a,T)
=\frac{k_BT}{2\pi}
\sum\limits_{l=0}^{\infty}
\left(1-\frac{\delta_{l0}}{2}\right)
\int_{0}^{\infty}k_{\bot}dk_{\bot}
\label{eq1} \\
&& \quad \phantom{aaaaaa}
\times\left\{
\ln\Big[1-r_{\|}^{2}(\xi_l,k_{\bot})e^{-2aq_l}\Big]
+
\ln\Big[1-r_{\bot}^{2}(\xi_l,k_{\bot})e^{-2aq_l}\Big]
\right\},
\nonumber
\end{eqnarray}
where the reflection coefficients for two independent
polarizations of electromagnetic field are defined as
\begin{equation}
r_{\|}(\xi_l,k_{\bot})=
\frac{{\ee}_lq_l-k_l}{{\ee}_lq_l+k_l},
\qquad
r_{\bot}(\xi_l,k_{\bot})=
\frac{k_l-q_l}{k_l+q_l}.
\label{eq2}
\end{equation}
\noindent
Here  $k_{\bot}$ is the magnitude of the
wave vector in the plane of plates,  
$\xi_l=2\pi k_BTl/\hbar$ are the Matsubara frequencies,
$k_B$  is the Boltzmann constant, 
$\varepsilon_l=\varepsilon(i\xi_l)$, and
\begin{equation}
q_l=\sqrt{\frac{\xi_l^2}{c^2}+k_{\bot}^2},
\quad
k_l=\sqrt{{\ee}(i\xi_l)\frac{\xi_l^2}{c^2}
+k_{\bot}^2}.
\label{eq3}
\end{equation}

The problems in the application of the Lifshitz theory  to real materials
discussed in the Introduction are closely connected with the values of the
reflection coefficients at zero Matsubara frequency. 
{For later use we discuss it for the various cases.}
\begin{itemize}
\item
For ideal metals\cite{32} it holds
\begin{equation}
r_{\|}(0,k_{\bot})=r_{\bot}(0,k_{\bot})=1.
\label{eq4}
\end{equation}
\item
For real metals described by the dielectric function of the Drude model,
\begin{equation}
\varepsilon(i\xi_l)=1+\frac{\omega_p^2}{\xi_l[\xi_l+\nu(T)]},
\label{eq5}
\end{equation}
where $\omega_p$ is the plasma frequency and $\nu(T)$ is the relaxation
parameter, it holds\cite{33}
\begin{equation}
r_{\|}(0,k_{\bot})=1,
\quad r_{\bot}(0,k_{\bot})=0.
\label{eq6}
\end{equation}
Eq.~(\ref{eq6}) results in the discontinuity between the cases
of ideal and real metals and leads to the violation of the Nernst
heat theorem for metallic plates having perfect crystal 
lattices.\cite{43,44,45,45b}

\item
For real metals described by the dielectric function of the plasma model,
\begin{equation}
\varepsilon(i\xi_l)=1+\frac{\omega_p^2}{\xi_l^2},
\label{eq7}
\end{equation}
\noindent
from Eq.~(\ref{eq2}) it follows:\cite{35,36}
\begin{equation}
r_{\|}(0,k_{\bot})=1,
\quad r_{\bot}(0,k_{\bot})=
\frac{\sqrt{c^2k_{\bot}^2+\omega_p^2}-ck_{\bot}}{\sqrt{c^2k_{\bot}^2
+\omega_p^2}+ck_{\bot}}.
\label{eq8}
\end{equation}
\noindent
Here, in the limit of ideal metals ($\omega_p\to\infty$) the continuity
is preserved because $r_{\bot}(0,k_{\bot})$ in
Eq.~(\ref{eq8}) goes to unity. The free energy (\ref{eq1}) calculated
with the permittivity (\ref{eq8}) is also consistent with thermodynamics.
\item
For dielectrics and semiconductors the dielectric permittivities  at the
imaginary Matsubara frequencies are given by the Ninham-Parsegian
representation,\cite{57,57a}
\begin{equation}
\varepsilon(i\xi_l)=1+\sum\limits_{j}\frac{C_j}{1+{\xi_l^2}/{\omega_j^2}},
\label{eq9}
\end{equation}
\noindent
where the parameters $C_j$ are the absorption strengths satisfying the
condition
\begin{equation}
\sum\limits_{j}C_j=\varepsilon_0-1
\label{eq10}
\end{equation}
\noindent
and $\omega_j$ are the characteristic absorption frequencies.
Here, the static dielectric permittivity 
$\varepsilon_0\equiv\varepsilon(0)$ is supposed to be finite.
Although Eq.~(\ref{eq10}) is an approximate one, it gives a very
accurate description for many materials.\cite{58} By the substitution
of Eq.~(\ref{eq9}) in Eq.~(\ref{eq2}) one arrives at
\begin{equation}
r_{\|}(0,k_{\bot})\equiv r_0=\frac{\varepsilon_0-1}{\varepsilon_0+1},
\quad r_{\bot}(0,k_{\bot})=0.
\label{eq11}
\end{equation}
\end{itemize}

\noindent
Note that the vanishing of the transverse reflection coefficient for
dielectrics at zero frequency in Eq.~(\ref{eq11}) has another meaning than
for the Drude metals in Eq.~(\ref{eq6}). For Drude metal the parallel
reflection coefficient is equal to the physical value for real photons
at normal incidence, i.e., to unity, and the transverse one vanishes
instead of taking unity, its physical value. This results in the violation
of the Nernst heat theorem for perfect crystal lattices. 
In the case of dielectrics both reflection
coefficients at zero frequency in Eq.~(\ref{eq11}) depart from the
physical value for real photons which is equal to 
$(\sqrt{\varepsilon_0}-1)/(\sqrt{\varepsilon_0}+1)$.
In this case, however, one of them is larger and the other one is smaller than 
the physical value. As we will see below, this leads to the
preservation of  Nernst's heat theorem confirming that Eq.~(\ref{eq9}),
{despite being approximate}, 
describes the material properties of dielectric and semiconductor
plates in a {thermodynamic} consistent way.

Now we derive the analytic representation for the Casimir free energy
in Eq.~(\ref{eq1}) at low temperatures. For convenience in calculations,
we introduce the dimensionless variables
\begin{equation}
\zeta_l=\frac{\xi_l}{\xi_c}=\frac{2a\xi_l}{c}=\tau l,
\qquad
y=2aq_l,
\label{eq12}
\end{equation}
\noindent
where $\xi_c=c/(2a)$ is the characteristic
frequency, $\tau=4\pi k_BaT/(\hbar c)$,
 and $q_l$ was defined in Eq.~(\ref{eq3}). Then
the Lifshitz formula (\ref{eq1}) takes the form
\begin{equation}
{\cal F}(a,T)=\frac{\hbar c\tau}{32\pi^2 a^3}
\sum\limits_{l=0}^{\infty}
\left(1-\frac{\delta_{l0}}{2}\right)
\int_{\zeta_l}^{\infty}dy\;f(\zeta_l,y)\,,
\label{eq13}
\end{equation}
where
\begin{eqnarray}
&&f(\zeta,y)=f_{\|}(\zeta,y)+f_{\bot}(\zeta,y),
\label{eq21}
\\
&&f_{\|,\bot}(\zeta,y)=y\ln\left[1-r_{\|,\bot}^2(\zeta,y)
e^{-y}\right],
\label{eq22}
\end{eqnarray}
and reflection coefficients (\ref{eq2}), 
in terms of variables (\ref{eq12}), being given by
\begin{equation}
r_{\|}(\zeta_l,y)=
\frac{{\ee}_ly-
\sqrt{y^2+\zeta_l^2({\ee}_l-1)}}{{\ee}_ly+
\sqrt{y^2+\zeta_l^2({\ee}_l-1)}}\,,
\quad
r_{\bot}(\zeta_l,y)=
\frac{\sqrt{y^2+\zeta_l^2({\ee}_l-1)}-
y}{\sqrt{y^2+\zeta_l^2({\ee}_l-1)}+y}\,.
\label{eq14}
\end{equation}

To separate the temperature independent contribution and
thermal correction in Eq.~(\ref{eq13})  we apply the Abel-Plana
formula,\cite{3,6}
\begin{equation}
\sum\limits_{l=0}^{\infty}
\left(1-\frac{\delta_{l0}}{2}\right)F(l)=
\int_{0}^{\infty}F(t)\;dt
+i\int_{0}^{\infty}dt\;
\frac{F(it)-F(-it)}{e^{2\pi t}-1}\,,
\label{eq15}
\end{equation}
where $F(z)$ is an analytic function in the right half-plane. 
Here, taking it as
\begin{equation}
F(x)=\int_{x}^{\infty}dy\,f(x,y)\,
\label{eq20}
\end{equation}
and using Eq.~(\ref{eq15}), we can identically rearrange Eq.~(\ref{eq13})
to the form
\begin{equation}
{\cal F}(a,T)=E(a)+\Delta{\cal F}(a,T)\,,
\label{eq16}
\end{equation}
where $E(a)$ is the energy of the van der Waals or Casimir
interaction at zero temperature,
\begin{equation}
E(a)=\frac{\hbar c}{32\pi^2a^3}
\int_{0}^{\infty}d\zeta\, \int_\zeta^\infty dy\, f(\zeta,y)\,,
\label{eq17}
\end{equation}
and $\Delta{\cal F}(a,T)$ is the thermal correction to this energy,
\begin{equation}
\Delta{\cal F}(a,T)=\frac{i\hbar c\tau}{32\pi^2a^3}
\int_{0}^{\infty}dt\; \frac{F(i\tau t)-
F(-i\tau t)}{e^{2\pi t}-1}\,.
\label{eq19}
\end{equation}
\noindent
Note that, in fact, Eq.~(\ref{eq19}) describes the dependence of the free
energy on the temperature arising from the dependence on temperature
of the Matsubara frequencies. Thus, $\Delta{\cal F}(a,T)$ in
(\ref{eq19}) coincides with the thermal correction to the energy,
defined as ${\cal F}(a,T)-{\cal F}(a,0)$, only for plate materials with
temperature independent properties.

The asymptotic expressions for the energy $E(a)$ at both short
and large separations are well known.\cite{6,29a,30} Below we find the
asymptotic expressions for the thermal correction (\ref{eq19})
under the conditions $\tau\ll 1$ and $\tau\gg 1$. 
Taking into account the 
definition of $\tau$ in Eq.~(\ref{eq12}), the asymptotic
expressions at $\tau\ll 1$  are applicable both at small and
large separations if the temperature is sufficiently low.

We begin with condition $\tau\ll 1$.
Let us substitute Eq.~(\ref{eq9}) in Eqs.~(\ref{eq21}) -- (\ref{eq14}), 
expand the
function $f(x,y)$ in powers of $x=\tau t$, and than integrate the
obtained expansion with respect to $y$ from $x$ to infinity in
order to find $F(x)$ in Eq.~(\ref{eq20}) and $F(ix)-F(-ix)$ in
Eq.~(\ref{eq19}). 

It is easy to check that $f_{\bot}(\zeta,y)$ does not contribute to
the leading, second, order in the expansion of $F(ix)-F(-ix)$ in
powers of $x$. Thus, we can restrict ourselves by the consideration of
the expansion
\begin{equation}
f_{\|}(x,y)=y\ln\left(1-r_0^2e^{-y}\right)+
\frac{2{\ee}_0\,r_0^2}{{\ee}_0+1}\,\frac{x^2}{y\left(e^y-r_0^2\right)}+
\frac{4\xi_c^2\,r_0^2}{({\ee}_0+1)\omega_1^2}\,\frac{yx^2}{e^y-r_0^2}+
\mbox{O}(x^3),
\label{eq23}
\end{equation}
\noindent
where $r_0$ was defined in Eq.~(\ref{eq11}). Note that for simplicity
we consider here only one oscillator in Eq.~(\ref{eq9}) and put 
$\omega_j=\omega_1$. The case of several oscillator modes can be
considered in an analogous way.

As a next step, we integrate Eq.~(\ref{eq23}) term by term according to 
Eq.~(\ref{eq20}), expand the partial results in powers of $x$ and sum up 
the obtained series. Thereby we obtain the following expressions:
\begin{eqnarray}
&&Z_1(x)\equiv\int_{x}^{\infty}y \, dy \,
\ln\left(1-r_0^2e^{-y}\right)
=
-\sum\limits_{n=1}^{\infty}\frac{(1+nx)e^{-nx}}{n^3}\,r_0^{2n}
\nonumber
\\
&&\phantom{aaaaa}=
-\mbox{Li}_3(r_0^2)-\frac{x^2}{2}\ln(1-r_0^2)+\mbox{O}(x^3)\,,
\label{eq25}
\\
&&Z_2(x)\equiv\frac{2{\ee}_0r_0^2x^2}{{\ee}_0+1}\int_{x}^{\infty}
\frac{dy}{y\left(e^y-r_0^2\right)} =
-\frac{2{\ee}_0x^2}{{\ee}_0+1}
\sum\limits_{n=1}^{\infty}r_0^{2n}\mbox{Ei}(-nx)\,,
\label{eq26}\\
&&Z_3(x)\equiv\frac{4r_0^2\xi_c^2x^2}{({\ee}_0+1)\omega_1^2}
\int_{x}^{\infty}
\frac{ydy}{e^y-r_0^2} =
\frac{4\xi_c^2x^2}{({\ee}_0+1)\omega_1^2}
\sum\limits_{n=1}^{\infty}r_0^{2n}
\frac{(1+nx)e^{-nx}}{n^2}
\nonumber
\\
&&\phantom{aaaaa}=\frac{4\xi_c^2}{({\ee}_0+1)\omega_1^2}
\left[x^2\,\mbox{Li}_2(r_0^2)-\frac{x^4}{2}
\frac{r_0^2}{1-r_0^2}+\mbox{O}(x^5)\right],
\label{eq28}
\end{eqnarray}
where Li${}_n(z)$ is the polylogarithm function and
Ei$(z)$ is the exponential integral function.

{}From these equations it follows
\begin{eqnarray}
&&Z_1(ix)-Z_1(-ix)=\mbox{O}(x^3)\,,
\qquad 
Z_3(ix)-Z_3(-ix)=\mbox{O}(x^5)\,,
\label{eq29}
\\
&&Z_2(ix)-Z_2(-ix)=2i\pi\,
\frac{{\ee}_0}{{\ee}_0+1}\,\frac{r_0^2}{1-r_0^2}\;x^2+\mbox{O}(x^3)\,,
\label{eq30}
\end{eqnarray}
\noindent
and, thus, $Z_1$ and $Z_3$ do not contribute to the leading
order in the expansion  of $F(ix)-F(-ix)$. The latter is determined by
$Z_2$ only. As a result, we arrive at
\begin{equation}
F(ix)-F(-ix)=i\pi\,
\frac{(\ee_0-1)^2}{2({\ee}_0+1)}\,x^2-i\alpha x^3+\mbox{O}(x^4),
\label{eq31}
\end{equation}
\noindent
where $r_0$ was substituted from  Eq.~(\ref{eq11}) and 
$\alpha$ was introduced for the still unknown real coefficient 
of the next to leading order resulting from $Z_1$ and $Z_2$ as well
as, possibly, from $f_{\bot}(\zeta,y)$. 
At this stage it is difficult to determine the value of
this coefficient because all powers in the expansion of $f(x,y)$
contribute to it. Remarkably, the two leading orders depend only on
the static dielectric permittivity ${\ee}_0$ and are not influenced
by the dependence of the dielectric permittivity on the frequency
contained in $Z_3$.

Substituting Eq.~(\ref{eq31}) in Eq.~(\ref{eq19}) and using
Eq.~(\ref{eq16}), we obtain
\begin{equation}
{\cal F}(a,T)=E(a)-\frac{\hbar c}{32\pi^2a^3}
\left[\frac{\zeta(3)}{8\pi^2}\,
\frac{({\ee}_0-1)^2}{{\ee}_0+1}\,\tau^3-C_4\tau^4+
\mbox{O}(\tau^5)\right],
\label{eq32}
\end{equation}
\noindent
where $C_4\equiv\alpha/240$ and $\zeta(z)$ is the Riemann
zeta function.

So far we have considered the free energy. The thermal pressure
is obtained as
\begin{equation}
P(a,T)=-\frac{\partial{\cal F}(a,T)}{\partial a}=
P_0(a)-\frac{\hbar c}{32\pi^2a^4}\left[
C_4\tau^4+\mbox{O}(\tau^5)\right],
\label{eq33}
\end{equation}
\noindent
where $P_0(a)=-\partial E(a)/\partial a$ is the Casimir pressure
at zero temperature.

In order to determine the value of the coefficient $C_4$ of the leading
term, we express the pressure directly through the Lifshitz
formula 
\begin{eqnarray}
\hspace{-.5cm}
P(a,T)=-\frac{\hbar c\tau}{32\pi^2a^4}
\sum\limits_{l=0}^{\infty}
\left(1-\frac{\delta_{l0}}{2}\right)
\!\! \int_{\zeta_l}^{\infty}\!\! y^2dy
\left[\frac{r_{\|}^{2}(\zeta_l,y)}{e^{y}-r_{\|}^{2}(\zeta_l,y)}+
\frac{r_{\bot}^{2}(\zeta_l,y)}{e^{y}-r_{\bot}^{2}(\zeta_l,y)}\right]\!.
\label{eq34}
\end{eqnarray}
\noindent 
Again, applying the Abel-Plana formula (\ref{eq15}), we represent the pressure
as follows,
\begin{equation}
P(a,T)=P_0(a)+\Delta P(a,T),
\label{eq35}
\end{equation}
\noindent
where the thermal correction to $P_0(a)$, the pressure 
at zero temperature, is
\begin{equation}
\Delta P(a,T)=-\frac{i\hbar c\tau}{32\pi^2a^4}
\int_{0}^{\infty}dt\;\frac{\Phi(i\tau t)-
\Phi(-i\tau t)}{e^{2\pi t}-1}
\label{eq36}
\end{equation}
\noindent
and the function $\Phi(x)$ is given by
\begin{equation}
\Phi(x)\equiv \Phi_{\|}(x)+\Phi_{\bot}(x)\,,
\qquad
\Phi_{\|,\bot}(x)=\int_{x}^{\infty}dy\,
\frac{y^2\,r_{\|,\bot}^{2}(x,y)}{e^{y}-r_{\|,\bot}^{2}(x,y)}\, .
\label{eq37}
\end{equation}

First, we determine the leading term of the expansion of
$\Phi_{\bot}(x)$ in powers of $x$. For this purpose,
let us introduce the new variable $v=y/x$ and note that 
 the reflection 
coefficient $r_{\bot}(x,v)$  depends
on $x$ only through the frequency dependence of ${\ee}$
given by Eq.~(\ref{eq9}).
Thus, we can rewrite and expand Eq.~(\ref{eq37}) as follows:
\begin{equation}
\Phi_{\bot}(x)=x^3\int_{1}^{\infty}dv\,
\frac{v^2\,r_{\bot}^{2}(x,v)}{e^{vx}-
r_{\bot}^{2}(x,v)}
=x^3\int_{1}^{\infty}dv\,
\frac{v^2\,r_{\bot}^{2}(v)}{1-r_{\bot}^{2}(v)}\,+\mbox{O}(x^4),
\label{eq39}
\end{equation}
\noindent
where, according to Eq.~(\ref{eq14}),
\begin{equation}
r_{\bot}(v)\equiv r_{\bot}(0,v)=\frac{\sqrt{v^2+{\ee}_0 -1}-
v}{\sqrt{v^2+{\ee}_0 -1}+v}\,.
\label{eq40}
\end{equation}
Integration in Eq.~(\ref{eq39}) with account of Eq.~(\ref{eq40})
results in
\begin{equation}
\Phi_{\bot}(x)=\left[1-
\frac{{\ee}_0(3-{\ee}_0)}{2\sqrt{\ee_0}}\right]\,\frac{x^3}{6}
+\mbox{O}(x^4)\,,
\label{eq41}
\end{equation}
\noindent
from  which it follows:
\begin{equation}
\Phi_{\bot}(ix)-\Phi_{\bot}(-ix)=-i\Big[1-
\frac{1}{2}\sqrt{\ee_0}\,(3-{\ee}_0)\Big]\frac{x^3}{3}
+\mbox{O}(x^5).
\label{eq42}
\end{equation}

The expansion of $\Phi_{\|}(x)$ from Eq.~(\ref{eq37}) in powers of $x$ is
somewhat more cumbersome. It can be performed in the following way.
As is seen from the second equality in Eq.~(\ref{eq29}), the dependence
of the dielectric permittivity on frequency  contributes to
$F(ix)-F(-ix)$ starting from only the 5th power in $x$. Bearing in mind
the connection between  free energy and pressure, we can conclude that
the dependence on the frequency contributes to 
$\Phi_{\|}(ix)-\Phi_{\|}(-ix)$ starting from the 4th order. We are
looking for the lowest (third) order expansion term of
$\Phi_{\|}(ix)-\Phi_{\|}(-ix)$. Because of this, it is permissible to
disregard the frequency dependence of $\ee$ and describe the dielectric
by its static dielectric permittivity.

To begin with, we identically rearrange $\Phi_{\|}(x)$
in Eq.~(\ref{eq37}) by subtracting and adding the two first expansion
terms of the function under the integral in powers of $x$,
\begin{eqnarray}
&&\Phi_{\|}(x)  = 
\int_{x}^{\infty}dy\left[y^2
\frac{r_{\|}^2(x,y)}{e^y-r_{\|}^2(x,y)}-
y^2\frac{r_0^2}{e^y-r_0^2}+
x^2\frac{2{\ee}_0}{{\ee}_0+1}\frac{r_0^2e^{-y}}{\left(1-r_0^2e^{-y}\right)^2}
\right]
\nonumber \\
&&\phantom{aaaaa}
+\int_{x}^{\infty}y^2dy\frac{r_0^2}{e^y-r_0^2}-
x^2\frac{2{\ee}_0}{{\ee}_0+1} \int_{x}^{\infty}dy
\frac{r_0^2e^{-y}}{\left(1-r_0^2e^{-y}\right)^2}\,,
\label{eq43}
\end{eqnarray}
and consider these three integrals separately.
The first integral in terms of the new variable $v=y/x$ reads
\begin{equation}
Q_1(x)\equiv x^3\int_{1}^{\infty}dv
\Bigg[v^2
\frac{r_{\|}^2(v)}{e^{vx}-r_{\|}^2(v)}-
v^2\frac{r_0^2}{e^{vx}-r_0^2}+
\frac{2{\ee}_0}{{\ee}_0+1}\frac{r_0^2e^{-vx}}{\left(1-
r_0^2e^{-vx}\right)^2}
\Bigg],
\label{eq44}
\end{equation}
where, in accordance with Eq.~(\ref{eq14}),
\begin{equation}
r_{\|}(v)\equiv r_{\|}(0,v)
=\frac{{\ee}_0v-\sqrt{v^2+{\ee}_0-1}}{{\ee}_0v+\sqrt{v^2+{\ee}_0-1}}\,.
\label{eq45}
\end{equation}

Expanding $Q_1(x)$ in powers of $x$
and explicitly calculating the remaining integrals for the lowest, 
third, power 
of $x$ results in
\begin{eqnarray}
Q_1(x)&=& x^3\int_{1}^{\infty}dv
\bigg[v^2
\frac{r_{\|}^2(v)}{1-r_{\|}^2(v)}-
v^2\frac{r_0^2}{1-r_0^2}+
\frac{2{\ee}_0}{{\ee}_0+1}\frac{r_0^2}{\left(1-r_0^2\right)^2}
\bigg]+\mbox{O}(x^4)
\label{eq47}
\\
&=&\frac{x^3}{6}\bigg[1-\frac{\sqrt{{\ee}_0}}{2}\left(1 + 3{\ee}_0
-2\,{\ee}_0^2 \right)+
\frac{2\,r_0^2}{1-r_0^2} -
\frac{12\,{\ee}_0}{{\ee}_0+1}\frac{r_0^2}{\left(1-r_0^2\right)^2}\bigg]\!
+\mbox{O}(x^4).
\nonumber
\end{eqnarray}
The second and third integrals on the right-hand side of Eq.~(\ref{eq43})
are simply determined  with the following result:
\begin{eqnarray}
&&Q_2(x)\equiv\int_{x}^{\infty}y^2dy\frac{r_0^2}{e^y-r_0^2}=
\sum\limits_{n=1}^{\infty}r_0^{2n}
\frac{(2+2nx+n^2x^2)e^{-nx}}{n^3}
\nonumber
\\
&&\phantom{aaaaa}=3\mbox{Li}_3(r_0^2)-\frac{x^3}{3}\frac{r_0^2}{1-r_0^2}
+\mbox{O}(x^4)\,,
\label{eq49}
\\
&&Q_3(x)\equiv
-\frac{2{\ee}_0x^2}{{\ee}_0+1} \int_{x}^{\infty}dy
\frac{r_0^2e^{-y}}{\left(1-r_0^2e^{-y}\right)^2}=
-\frac{2{\ee}_0x^2}{{\ee}_0+1}
\frac{r_0^2e^{-x}}{1-r_0^2e^{-x}}
\nonumber
\\
&&\phantom{aaaaa}=-\frac{2{\ee}_0}{{\ee}_0+1}\left[x^2
\frac{r_0^2}{1-r_0^2}-x^3
\frac{r_0^2}{(1-r_0^2)^2}\right]
+\mbox{O}(x^4)\,.
\label{eq51}
\end{eqnarray}

Substituting Eqs.~(\ref{eq47}), (\ref{eq49}) and (\ref{eq51})
into  $\Phi_{\|}(x)=Q_1(x)+Q_2(x)+Q_3(x)$,
we arrive at
\begin{equation}
\Phi_{\|}(ix)-\Phi_{\|}(-ix)=-i\left[1+
\frac{\sqrt{{\ee}_0}}{2}\left(2{\ee}_0^2-3{\ee}_0-1\right)\right]
\frac{x^3}{3}
+\mbox{O}(x^4).
\label{eq52}
\end{equation}
\noindent
Then, by summing Eqs.~(\ref{eq42}) and (\ref{eq52}), the result is obtained
\begin{equation}
\Phi(ix)-\Phi(-ix)=-i\left(2+{\ee}_0^{5/2}-{\ee}_0^{3/2}-
2\sqrt{{\ee}_0}\right)
\frac{x^3}{3}
+\mbox{O}(x^4).
\label{eq53}
\end{equation}

Now we substitute Eq.~(\ref{eq53}) in Eq.~(\ref{eq36}) and perform
integration. Finally, from Eq.~(\ref{eq35}) the desired expression for
the Casimir pressure is derived
\begin{equation}
P(a,T)=P_0(a)-\frac{\hbar c}{32\pi^2a^4}
\bigg[\frac{\left(\sqrt{{\ee}_0}-1\right)\big(
{\ee}_0^2+{\ee}_0^{3/2}-2\big)}{720}\,\tau^4 +\mbox{O}(\tau^5)\bigg].
\label{eq54}
\end{equation}
By comparison with Eq.~(\ref{eq33}) the explicit form of the
coefficient $C_4$ is found as
\begin{equation}
C_4={\big(\sqrt{{\ee}_0}-1\big)\big(
{\ee}_0^2+{\ee}_0^{3/2}-2\big)}/{720}
\label{eq55}
\end{equation}
\noindent
and, thus, both two first perturbation orders in the expansion for the
free energy (\ref{eq32}) are determined.

Equations (\ref{eq32}), (\ref{eq54}) and (\ref{eq55}) solve the
fundamental problem of the thermodynamic consistency of the Lifshitz
theory in the case of two dielectric plates. From Eqs.~(\ref{eq32}) and
(\ref{eq55}) the entropy of the van der Waals and Casimir interaction
between plates takes the form
\begin{eqnarray}
&&S(a,T)=-\frac{\partial{\cal F}(a,T)}{\partial T}=
\frac{3k_B\zeta(3)({\ee}_0-1)^2}{64\pi^3a^2({\ee}_0+1)}\tau^2
\label{eq56} \\
&&\phantom{aaaaaa}\times
\bigg[1-
\frac{2\pi^2({\ee}_0+1)\big({\ee}_0^{3/2}+2{\ee}_0+
2\sqrt{{\ee}_0}+2\big)}{135\zeta(3)(\sqrt{{\ee}_0}+1)^2}
\,\tau+\mbox{O}(\tau^2)\bigg].
\nonumber
\end{eqnarray}

As is seen from Eq.~(\ref{eq56}), in the limit $\tau\to 0$ ($T\to 0$) the
lower order contributions to the entropy are of the second and the third 
powers in the small parameter $\tau$. Thus, the entropy vanishes when the 
temperature goes to zero as it must be in accordance with the third law
of thermodynamics (the Nernst heat theorem). 

A similar behavior was
obtained for ideal metals\cite{32,59,60} and for real metals described by
the plasma model.\cite{36,43} For example, in the case of plates made of ideal
metal the entropy at low temperatures is given by\cite{32,59,60}
\begin{equation}
S(a,T)=\frac{3k_B\zeta(3)}{32\pi^3a^2}\,\tau^2\left[1-
\frac{2\pi^2}{135\zeta(3)}\,\tau+\mbox{O}(\tau^2)\right].
\label{eq57}
\end{equation}
\noindent
Note, however, that the expansion coefficients in Eq.~(\ref{eq57})
cannot be obtained as a straightforward limit 
$|{\ee}_0|\to\infty$ in Eq.~(\ref{eq56}) and the above equations for
the free energy and pressure. The mathematical reason is that it is
impermissible to interchange the limiting transitions $\tau\to 0$
and $|{\ee}_0|\to\infty$ in the power expansions of functions depending
on ${\ee}_0$ as a parameter.

Remarkably, the low-temperature behavior of the free energy, pressure 
and entropy of nonpolar dielectrics in Eqs.~(\ref{eq32}), (\ref{eq54})
and (\ref{eq56}) is universal, i.e., is determined only by the
static dielectric permittivity. The absorption bands included in 
Eq.~(\ref{eq9}) do not influence the low-temperature behavior.
A more simple derivation of the results (\ref{eq32}), 
(\ref{eq54})--(\ref{eq56}) for dielectrics with constant $\ee$ is
contained in Ref.~\refcite{61}. As was demonstrated above, all these results 
remain unchanged if the dependence of dielectric permittivity on
frequency is taken into account.

In Ref.~\refcite{48} more general results were obtained related to 
two dissimilar
dielectric plates with dielectric permittivities
$\varepsilon^{(1)}(\omega)$ and $\varepsilon^{(2)}(\omega)$.
For brevity here we present only the final expressions for the
low-temperature behavior of the Casimir free energy, pressure 
and entropy between dissimilar plates. They are as follows:\cite{48}
\begin{eqnarray}
&&{\cal F}(a,T)=E(a)-\frac{\hbar c}{32\pi^2a^3}\,\tau^3 
\nonumber \\
&&~~ \times
\Bigg[\frac{\zeta(3)}{8\pi^2}\;
\frac{\ezf +\ezs + 2\ezf \ezs}{(\ezf +1)(\ezs +1)}\;
\frac{(\ezf -1)(\ezs -1)}{(\ezf +\ezs)}
-C_4\,\tau+\mbox{O}(\tau^2)\Bigg],
\nonumber \\
&&P(a,T)=P_0(a)-\frac{\hbar c}{32\pi^2a^4}\left[C_4\tau^4+
\mbox{O}(\tau^5)\right],
\phantom{\frac{(\ezf -1)(\ezs -1)}{8\pi^2(\ezf +\ezs)}}
\label{eq58} \\
&&S(a,T)=\frac{k_B}{2\pi a^2}\,\tau^2 
\nonumber \\
&&~~ \times
\left[\frac{3\zeta(3)}{32\pi^2}\;
\frac{\ezf+\ezs+2\ezf\ezs}{(\ezf+1)(\ezs+1)}\;
\frac{(\ezf-1)(\ezs-1)}{(\ezf+\ezs)}-
C_4\,\tau+\mbox{O}(\tau^2)\right].
\nonumber
\end{eqnarray}
\noindent
Here $\varepsilon_0^{(1,2)}\equiv\varepsilon^{(1,2)}(0)$ and the
coefficient $C_4$ is given by\cite{48}
\begin{eqnarray}
&&
C_4=\frac{1}{720}
\left\{
2+
\frac{1}{\Big(\sqrt{\ezf}+
\sqrt{\ezs}\,\Big)\left(\ezf +\ezs \right)^2}
\right.
\label{eq59} \\
&&\phantom{a}
\times\left[
\vphantom{\frac{3\big({\ezf} {\ezs}\big)^2\big(\ezf -1\big)\big(\ezs - 1\big)}
{\Big(\sqrt{\ezf} -\sqrt{\ezs}\,\Big)\sqrt{\ezf +\ezs}}}
-\left({\ezf +\ezs}\right)^2
\Big(2\ezf +2\ezs +\sqrt{\ezf \ezs}-\ezf \ezs\Big)
\right.
\nonumber \\
&&\phantom{aa}
+\ezf \ezs \sqrt{\ezf \ezs}
\left(5\ezf \ezs -3\ezf -3\ezs+1\right)
\nonumber \\
&&\phantom{aa}
+\sqrt{\ezf \ezs}\Big(\sqrt{\ezf}-\sqrt{\ezs}\,\Big)^2
\Big(\ezf \ezs \sqrt{\ezf \ezs}-\sqrt{\ezf \ezs}-\ezf -\ezs\Big) 
\nonumber \\
&&\phantom{aa}
\left.\left.
-
\frac{3\big({\ezf} {\ezs}\big)^2\big(\ezf -1\big)\big(\ezs - 1\big)}
{\Big(\sqrt{\ezf} -\sqrt{\ezs}\,\Big)\sqrt{\ezf +\ezs}}\;
\mbox{Artanh}\frac{\sqrt{\ezf +\ezs}\Big(\sqrt{\ezf} -
\sqrt{\ezs}\,\Big)}{\sqrt{\ezf \ezs}-\ezf -\ezs}\right]\right\}\, .
\nonumber
\end{eqnarray}
\noindent
It is easily seen that in the limit $\ezf=\ezs={\ee}_0$ equations 
(\ref{eq58}), (\ref{eq59}) coincide with equations 
(\ref{eq32}), (\ref{eq54})--(\ref{eq56}) having obtained above. Note that in the application
region of low-temperature asymptotic expressions the entropy
of the Casimir interaction between dielectric plates is nonnegative.
\begin{figure}[t]
\vspace*{-2.2cm}
\centerline{\psfig{file=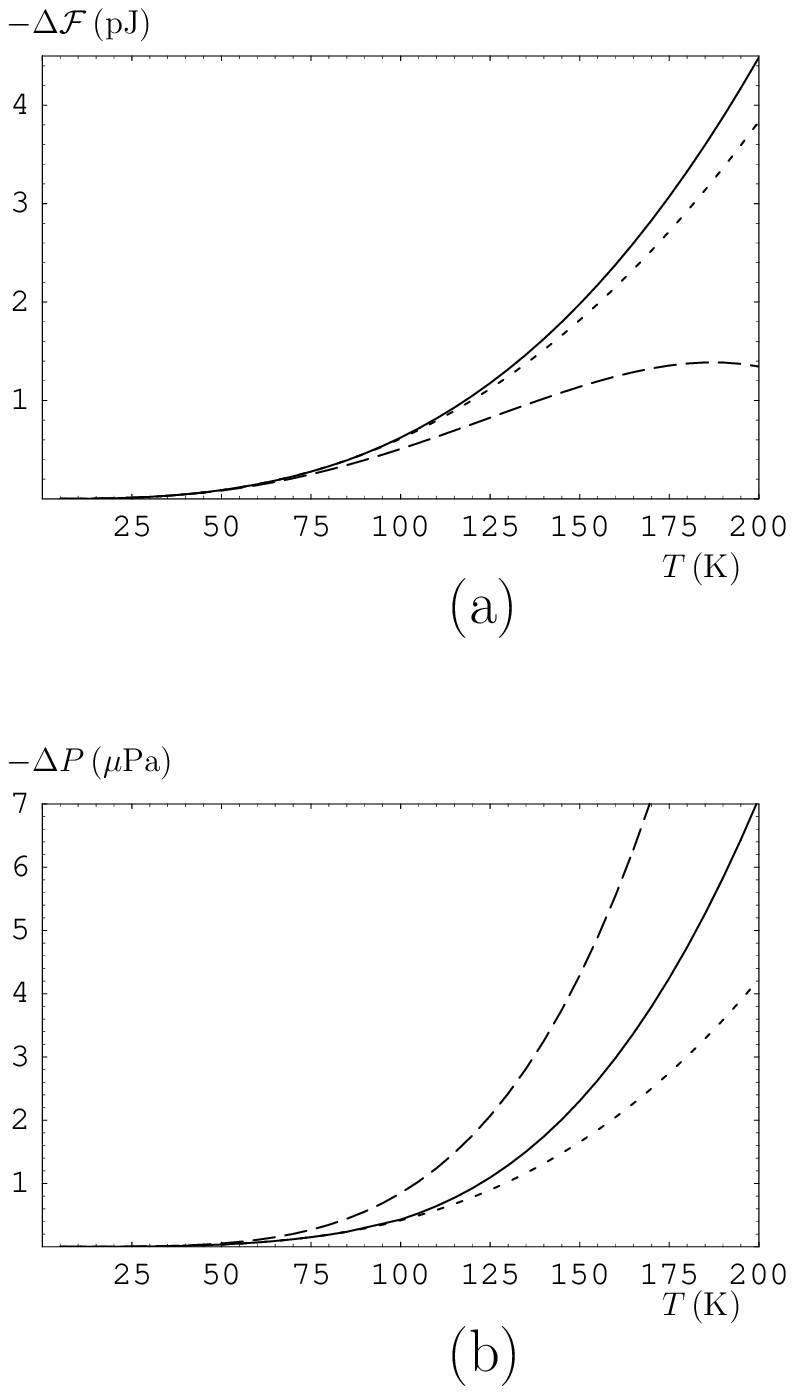,width=20cm}}
\vspace*{-11.5cm}
\caption{\label{fg1}
Magnitudes of the thermal corrections to the energy (a)
and pressure (b) in configuration of two plates, one made of
Si and another one of $\mbox{SiO}_2$,  at a separation
$a=400\,$nm as a function of temperature calculated by the
use of different approaches: by the Lifshitz formula and 
tabulated optical data (solid lines), by the Lifshitz formula 
and static dielectric permittivities (short-dashed lines),
by the asymptotic expressions in Eqs.~(\ref{eq58}) and 
(\ref{eq59}) (long-dashed lines).}
\end{figure}

The obtained analytic behavior of the free energy, pressure and entropy 
at low temperatures can be compared with the results of numerical
computations using the Lifshitz formula. Dielectric properties of the
plates can be described by the static dielectric permittivity or more
precisely using the optical tabulated data for the complex index of
refraction. As an example, in Fig.~1 we present the thermal corrections
to the Casimir energy (a) and pressure (b) at a separation $a=400\,$nm
as functions of temperature in the configuration of two dissimilar
plates made of high-resistivity Si and SiO${}_2$. The dielectric
permittivities of both materials along the imaginary frequency axis 
were computed in Ref.~\refcite{62} using the optical data of 
Ref.~\refcite{63}. 
The precise thermal corrections computed by {taking into account} these 
permittivities are shown by the solid lines and corrections
computed by our analytical asymptotic expressions are shown by the
long-dashed lines. Short-dashed lines indicate the results computed
by the Lifshitz formula with constant dielectric permittivities
of Si  and SiO${}_2$ equal to $\ezf=11.67$ and $\ezs=3.84$,
respectively. As is seen in Fig.~1a,b, at $T<60\,$K the results
obtained using the analytical asymptotic expressions practically
coincide with the solid lines computed using the tabulated optical
data for the materials of the plates.

Now we return to the case of two similar dielectric plates and
consider the asymptotic expressions under the condition $\tau\gg 1$,
i.e., at high temperatures (large separations). It is well
known\cite{29a,30,63a} that in this case the approximation of static
dielectric permittivity works good and the main contribution
is given by the zero-frequency term of the Lifshitz formula
(\ref{eq13})
\begin{equation}
{\cal F}(a,T)=\frac{\hbar c\tau}{64\pi^2a^3}
\int_{0}^{\infty}ydy\ln\left(1-r_0^2e^{-y}\right)
\label{eq59a}
\end{equation}
\noindent
(the other terms being exponentially small). Performing the
integration in Eq.~(\ref{eq59a}) we obtain
\begin{equation}
{\cal F}(a,T)=-\frac{k_B T}{16\pi a^2}\;
\mbox{Li}_3\left(r_0^2\right).
\label{eq59b}
\end{equation}

In a similar manner for the Casimir pressure and entropy at
$\tau\gg 1$ it follows
\begin{equation}
P(a,T)=-\frac{k_B T}{8\pi a^3}\;
\mbox{Li}_3\left(r_0^2\right),
\qquad
S(a,T)=\frac{k_B}{16\pi a^2}\;
\mbox{Li}_3\left(r_0^2\right).
\label{eq59c}
\end{equation}
\noindent
Equations (\ref{eq59b}) and (\ref{eq59c}) are simply generalized\cite{48}
for the case of two dissimilar dielectric plates by performing the
replacement $r_0^2\to r_0^{(1)}r_0^{(2)}$, where $r_0^{(1,2)}$ are
defined by Eq.~(\ref{eq11}) with the static dielectric permittivities
of dissimilar plates $\ee_0^{(1,2)}$.

\section{IS THE DC CONDUCTIVITY RELATED TO THE CASIMIR INTERACTION
BETWEEN DIELECTRICS?}

As was discussed in the previous section, {the zero-frequency term
in formula (\ref{eq13}), i.e., the contribution with $l=0$, }
is of prime importance and determines many of the basic properties of 
the Casimir interaction. In the above consideration we have described
dielectric materials by Eq.~(\ref{eq9}) with finite static dielectric
permittivity $\ez$. This resulted in Eq.~(\ref{eq11}) where one
reflection coefficient at zero frequency is larger and the other one
is smaller than the physical value for real photons at normal
incidence. {However, in the Lifshitz theory, the departure} of both 
coefficients from their physical values is coordinated in such a way 
that the Nernst heat theorem remains valid.

It is common knowledge that at nonzero temperatures dielectric
materials possess a negligibly small but not equal to zero dc conductivity.
{}From physical intuition it is reasonable to expect that the 
influence of this conductivity on
the van der Waals and Casimir forces should be also negligible.
In Ref.~\refcite{46} it was shown that, on the contrary, the inclusion of
small dielectric dc conductivity in the model of dielectric response
leads to a large effect in dispersion forces. This raises the question if  
dc conductivity is related to dispersion forces {or if, on the contrary,}
the zero-frequency contribution should be understood not literally
but as an analytic continuation from the region of high frequencies
determining the physical phenomenon of dispersion forces.

To illustrate this problem in more details, we consider the asymptotic
behavior of the free energy and entropy at low temperature with
included dc conductivity. What this means is that, instead of the
dielectric permittivity $\varepsilon(i\xi_l)$ given in Eq.~(\ref{eq9}),
one uses\cite{46,46a}
\begin{equation}
\tilde{\ee}(i\xi_l)
={\ee}(i\xi_l)+\frac{4\pi\sigma_0}{\xi_l}=
{\ee}_l(i\xi)+\frac{\beta(T)}{l},
\label{eq60}
\end{equation}
\noindent
where $\sigma_0$ is the dc conductivity of the plate
material and $\beta(T)=2\hbar\sigma_0/(k_BT)$.
The conductivity of 
dielectrics depends on  temperature as 
$\sigma_0\sim\exp(-b/T)$ where $b$ is
determined by the energy gap $\Delta\!$ {which 
differs} for different materials.\cite{64}
The smallness of the dc conductivity of dielectrics can be 
illustrated\cite{65}
by the example of SiO${}_2$ where at $T=300\,$K
it holds $\beta\sim 10^{-12}$. Thus, the role of the dc conductivity
is really negligible for all $l\geq 1$. 
In addition, $\beta(T)$ quickly decreases with decrease
of $T$ and, as a consequence, remains negligible at any $T$.
 In spite of this, the substitution of Eq.~(\ref{eq60}) into
the reflection coefficients (\ref{eq14}) leads to 
different result than 
in Eq.~(\ref{eq11}):
\begin{equation}   
\tilde{r}_{\|}(0,y)=1,
\quad
\tilde{r}_{\|}(0,y)=0.
\label{eq61}
\end{equation}
\noindent
Equation (\ref{eq61}) is in some analogy to Eq.~(\ref{eq6}), obtained for 
metals described by the Drude model, which leads to the violation of
the Nernst heat theorem in the case of perfect crystal lattices.
 
Now we substitute the dielectric permittivity 
$\tilde{\ee}_l\equiv\tilde{\ee}(\xi_l)$ in
Eqs.~(\ref{eq13}) -- (\ref{eq14}) instead of $\ee_l$ and find the Casimir
free energy $\tilde{\cal F}(a,T)$ with included dc conductivity.
For convenience, we separate the zero-frequency term,
subtract and add the
usual zero-frequency contribution for dielectric without dc
conductivity. The result is
\begin{eqnarray}
\tilde{\cal F}(a,T)
&=& \frac{k_BT}{16\pi a^2}
\int_{0}^{\infty}ydy\Big[
\ln\left(1-e^{-y}\right)
-\ln\left(1-r_0^2e^{-y}\right)\Big]
\nonumber \\
&
+&\frac{k_BT}{16\pi a^2}
\int_{0}^{\infty}ydy
\ln\left(1-r_0^2e^{-y}\right)
\label{eq62}\\
&
+&\frac{k_BT}{8\pi a^2}\sum\limits_{l=1}^{\infty}
\int_{\zeta_l}^{\infty}ydy\left\{
\ln\Big[1-\tilde{r}_{\|}^{2}(\zeta_l,y) e^{-y}\Big]+
\ln\Big[1-\tilde{r}_{\bot}^{2}(\zeta_l,y)e^{-y}\Big]
\right\},
\nonumber
\end{eqnarray}
\noindent
where $r_0$ was defined in Eq.~(\ref{eq11}).
Let us expand the last, third, integral on the right-hand side
of Eq.~(\ref{eq62}) in powers of the small parameters
$\beta/l$. Then, we combine the zero-order contribution in
this expansion with the second integral on the right-hand side
of Eq.~(\ref{eq62}) and obtain the Casimir free energy
${\cal F}(a,T)$ calculated with the dielectric permittivities
${\ee}_l$. The first integral on the 
right-hand side of Eq.~(\ref{eq62}) is calculated explicitly. 
Then, Eq.~(\ref{eq62}) can be rewritten as
\begin{equation}
\tilde{\cal F}(a,T)= {\cal F}(a,T)
-\frac{k_BT}{16\pi a^2}\left[\zeta(3)-
\mbox{Li}_3\left(r_0^2\right)\right]+R(a,T).
\label{eq63}
\end{equation}
\noindent
Here, $R(a,T)$ is of order $O(\beta/l)$. 
It represents the first and higher-order contributions in the
expansion of the third integral on the right-hand side
of Eq.~(\ref{eq62}) in powers of $\beta/l$.
Restricting its explicit form  to the first order contribution 
we get
\begin{eqnarray}
&&R(a,T)=R_1(a,T)+\mbox{\large$O$}\big[\left(
{\beta/l}\right)^2\big]\,,
\label{eq64}
\\
&&R_1(a,T)
=\frac{k_BT}{4\pi a^2}
\sum\limits_{l=1}^{\infty}
\frac{\beta}{l}\int_{\zeta_l}^{\infty}\!\!
\frac{dy\,y^2e^{-y}}{\sqrt{y^2+\zeta_l^2({\ee}_l-1)}}
\left\{\frac{(2-{\ee}_l)\,\zeta_l^2- 2y^2}
{\big[{\sqrt{y^2+\zeta_l^2({\ee}_l-1)}+\ee}_ly\,\big]^2}\,\,
\right.
\nonumber \\
&&\phantom{aaaa}\left.
\times \frac{r_{\|}(\zeta_l,y)}{1-
r_{\|}^2(\zeta_l,y)e^{-y}}
-\frac{\zeta_l^2}{\big[\sqrt{y^2+\zeta_l^2({\ee}_l-1)}+y\,\big]^2}\,\,
\frac{r_{\bot}(\zeta_l,y)}{1-r_{\bot}^2(\zeta_l,y)e^{-y}}
\right\}.
\label{eq65}
\end{eqnarray}

Calculating the entropy by the first equality in Eq.~(\ref{eq56}),
we arrive at
\begin{equation}
\tilde{S}(a,T)=S(a,T)+\frac{k_B}{16\pi a^2}\left[\zeta(3)-
\mbox{Li}_3\left(r_0^2\right)\right]-
\frac{\partial R(a,T)}{\partial T},
\label{eq66}
\end{equation}
\noindent
where $S(a,T)$ is the entropy for the plates with the dielectric
permittivity $\ee_l$ given by Eq.~(\ref{eq56}).

Let us now show that the quantity $R(a,T)$ exponentially goes to zero
with the decrease of $T$.  
First we consider only the integral in
Eq.~(\ref{eq65}), expand the integrated function in powers of
$\tau$ (we recall that $\zeta_l=\tau l$), restrict
ourselves to the main contribution, resulting for $\tau =0$, and rearrange it 
appropriately:
\begin{eqnarray}
&&-2\int_{\zeta_l}^{\infty}dy\frac{ye^{-y}}{(\ez +1)^2}\,
\frac{\rz}{1-r_0^2 e^{-y}}
= -\frac{2}{({\ee}_0^2-1)}
\int_{\zeta_l}^{\infty}dy\,y
\frac{r_0^2  e^{-y}}{1-r_0^2 e^{-y}}
\nonumber \\
&&\qquad
=
-\frac{2}{({\ee}_0^2-1)}\sum\limits_{n=1}^{\infty}
r_0^{2n}
\int_{\zeta_l}^{\infty}dy\,ye^{-ny}
=
-\frac{2}{({\ee}_0^2-1)}\sum\limits_{n=1}^{\infty}
r_0^{2n}
\frac{1+n\zeta_l}{n^2}e^{-n\zeta_l}.
\label{eq68}
\end{eqnarray}
Substituting this 
in Eq.~(\ref{eq65}), we find
\begin{eqnarray}
&&R_1(a,T)=
-\frac{k_BT\beta}{4\pi a^2\left({\ee}_0^2-1\right)}
\sum\limits_{n=1}^{\infty}
\frac{r_0^{2n}}{n^2}
\left(\sum\limits_{l=1}^{\infty}\frac{e^{-n\tau l}}{l}
+n\tau\sum\limits_{l=1}^{\infty}e^{-n\tau l}\right)
\nonumber \\
&&\phantom{aaaaaaa}
=-\frac{k_BT\beta}{4\pi a^2\left({\ee}_0^2-1\right)}
\sum\limits_{n=1}^{\infty}
\frac{r_0^{2n}}{n^2}
\left[-\ln\left(1-e^{-n\tau}\right)+
\frac{n\tau}{e^{n\tau}-1}\right]
\nonumber \\
&&\phantom{aaaaaaa}
=\frac{k_B\,\mbox{Li}_2\left(r_0^2\right)}{2\pi a^2\left({\ee}_0^2
-1\right)}\;
T\beta\,\ln\tau+T\beta\,\mbox{O}(\tau^0)\,.
\label{eq71}
\end{eqnarray}
Here, the last line is obtained by using the equality
\begin{equation}
-\ln\left(1-e^{-n\tau}\right)+
\frac{n\tau}{e^{n\tau}-1}=-\ln\tau+1-\ln n+\mbox{O}(\tau^2)\,,
\label{eq70}
\end{equation}
\noindent
substituting only its leading term and observing the definition 
of the integral logarithm,
$\mbox{Li}_2\left(z\right) = (1/2)\sum\limits_{n=1}^{\infty}
{z^{n}}/{n^2}$.
Taking into account that 
$\beta\sim(1/T)\exp(-b/T)$, we get the
conclusion that the temperature dependence of $R_1(a,T)$ is
given by 
\begin{equation}
R_1(a,T)\sim e^{-b/T}\ln T.
\label{eq72}
\end{equation}
\noindent
Thus, both $R_1(a,T)$ and its derivative with respect to $T$
in Eqs.~(\ref{eq63}) and (\ref{eq66}) go to zero.
The terms of the second and higher powers in $\beta$ in Eq.~(\ref{eq64})
go to zero even faster than $R_1$ when $T\to 0$.

Finally, in the limit $T\to 0$ from Eq.~(\ref{eq66}) it follows
\begin{equation}
\tilde{S}(a,0)=\frac{k_B}{16\pi a^2}\left[\zeta(3)-
\mbox{Li}_3\left(r_0^2\right)\right]>0.
\label{eq73}
\end{equation}
\noindent
The right-hand side of this equation depends on the parameter of the
system under consideration (the separation distance $a$) and implies
a violation of the Nernst heat theorem. An analogous result was
obtained\cite{48} in the case of two dissimilar dielectrics.

The violation of the Nernst heat theorem in the Casimir interaction
for dielectrics originates from the inclusion of the dc conductivity in
the model of dielectric response. This violation is, however, of different
nature than the one discussed above in the case of Drude metals. In the case
of dielectrics the entropy at zero temperature is positive but in the
case of Drude metals it is negative. In the case of metals the violation is
caused by the vanishing contribution from the transverse electric mode
at zero frequency whereas the other reflection coefficient takes the
physical value 1 [see Eq.~(\ref{eq6})]. For dielectrics the situation
is quite opposite. 
In this case the transverse reflection coefficient at zero
frequency is always equal to zero [compare  Eqs.~(\ref{eq11}) and 
(\ref{eq61}) in the absence and in the presence of the contribution from
dc conductivity]. Here the violation occurs due to the unity value of
the parallel reflection coefficient at zero frequency in 
Eqs.~(\ref{eq61}) which departs from the value $\rz=(\ez-1)/(\ez+1)$
coordinated with the zero value of the transverse coefficient
in Eq.~(\ref{eq11}).

One can conclude that the dc conductivity of a dielectric is not related
to the nature of the van der Waals and Casimir forces and must not be 
included in the model of dielectric response. {Ignoring this 
rule} results in a violation of thermodynamics. Physically it is
amply clear that there is no fluctuating field of zero frequency and
that for such high-frequency phenomena as the van der Waals and Casimir 
forces the low-frequency behavior should be obtained by  analytic
continuation from the region of high frequencies. This permits to
conclude that the correct {procedure consists in the substitution} 
of the finite
static dielectric permittivities into the zero-frequency term of
the Lifshitz formula, as Lifshitz and his collaborators really
did.\cite{29,29a,30}

\section{THERMAL CASIMIR FORCE BETWEEN DIELECTRIC AND METAL PLATES}

The Casimir interaction between metal and dielectric plates suggests 
the interesting opportunity to verify the thermodynamic consistency
of the Lifshitz theory with different models of the dielectric
response.  This configuration was first investigated in Ref.~\refcite{49}
where it was proved that the Casimir entropy is in accordance with
the demands of 
the Nernst heat theorem if the static permittivity of the dielectric
plate is finite. In Ref.~\refcite{49}, however, only the first leading terms 
in the low-temperature asymptotic expressions for the free energy and
entropy were obtained and the Casimir pressure was derived only in
the dilute approximation. Here we derive the more precise low-temperature
behavior for the Casimir free energy, pressure and entropy in the 
configuration of one plate made of ideal metal and another plate made
of dielectric with any finite static dielectric permittivity.

For the configuration of metal and dielectric plates the Lifshitz
formula takes the form\cite{65a} analogical to Eq.~(\ref{eq13})
\begin{eqnarray}
&&{\cal F}(a,T)=\frac{\hbar c\tau}{32\pi^2 a^3}
\sum\limits_{l=0}^{\infty}
\left(1-\frac{\delta_{l0}}{2}\right)
\int_{\zeta_l}^{\infty}ydy
\label{eq74} \\
&&\phantom{aaaa}
\times\left\{
\ln\Big[1-r_{\|}^{M}(\zeta_l,y)\,r_{\|}^{D}(\zeta_l,y)\,e^{-y}\Big]+
\ln\Big[1-r_{\bot}^{M}(\zeta_l,y)\,r_{\bot}^{D}(\zeta_l,y)\,e^{-y}\Big]
\right\}.
\nonumber
\end{eqnarray}
\noindent
Here the reflection coefficients $r_{\|,\bot}^{M,D}$ for metal and
dielectric, respectively, are given by Eqs.~(\ref{eq14}) where
$\ee_l$ should be changed for $\ee_l^{M,D}=\ee^{M,D}(i\xi_l)$.

For an ideal metal $r_{\|,\bot}^{M}(\zeta_l,y)=1$ and Eq.~(\ref{eq74})
takes the more simple form
\begin{eqnarray}
{\cal F}(a,T)&=&\frac{\hbar c\tau}{32\pi^2 a^3}
\sum\limits_{l=0}^{\infty}
\left(1-\frac{\delta_{l0}}{2}\right)
\int_{\zeta_l}^{\infty}ydy
\label{eq75} \\
&
\times&\left\{
\ln\left[1-r_{\|}(\zeta_l,y)
e^{-y}\right]
+\
\ln\left[1-r_{\bot}(\zeta_l,y)
e^{-y}\right]\right\}
\nonumber
\end{eqnarray}
\noindent
(here and below we omit the index $D$ near the reflection coefficient 
and permittivity of a dielectric plate). We admit that the dielectric
permittivity calculated at Matsubara frequencies $\ee_l\equiv\ez$,
i.e., is equal to its static value and find the asymptotic behavior
of Eq.~(\ref{eq75}) at small $\tau$. [In analogy with Sec.~2 it is
possible to prove that the deviations of $\ee(i\xi_l)$ from $\ez$
at high frequencies do not influence the low-temperature behavior
of the Casimir free energy, pressure and entropy. 
It can be shown also that the results of this section are valid not
only for ideal metal plate but for plate made of real metal as well.]
The free energy
(\ref{eq75}) can be represented by Eqs.~(\ref{eq20}) --
(\ref{eq19}), with the function $f(\zeta,y)$  replaced by
\begin{eqnarray}
&&{\hat f}(\zeta,y)=y\ln\left[1-r_{\|}(\zeta,y)e^{-y}\right]+
y\ln\left[1-r_{\bot}(\zeta,y)e^{-y}\right]
\nonumber \\
&&\phantom{aaaaa}
\equiv {\hat f}_{\|}(\zeta,y)+{\hat f}_{\bot}(\zeta,y).
\label{eq76}
\end{eqnarray}
\noindent
In the case of one dielectric and one metal plate both ${\hat f}_{\|}$ and
${\hat f}_{\bot}$ contribute to $F(ix)-F(-ix)$. The expansion
of ${\hat f}(x,y)$ in powers of $x$ takes the form
\begin{equation}
{\hat f}(x,y)=y\ln(1-\rz e^{-y})-\bigg(\frac{\ez-1}{4y}e^{-y}-
\frac{\ez}{\ez+1}\sum\limits_{n=1}^{\infty}r_0^n
\frac{e^{-ny}}{y}\bigg)\,x^2+\mbox{O}(x^3).
\label{eq77}
\end{equation}
\noindent
Now we integrate Eq.~(\ref{eq77}) in accordance with Eq.~(\ref{eq20})
to find the function $F(x)$. The integral of the first term on the
right-hand side of Eq.~(\ref{eq77}) is evaluated using the new
variable $v=y-x$:
\begin{equation}
\int_{x}^{\infty}ydy\ln(1-\rz e^{-y})=
\int_{0}^{\infty}vdv\ln(1-\rz e^{-v})+\mbox{O}(x^2),
\label{eq78}
\end{equation}
\noindent
where the coefficient near the first-order contribution in $x$
vanishes. As a result, this term could contribute to $F(ix)-F(-ix)$
only starting from the third expansion order. The integrals of the
second-order terms  on the
right-hand side of Eq.~(\ref{eq77}) are simply calculated using
the formulas
\begin{equation}
\int_{x}^{\infty}dy\frac{e^{-y}}{y}=-\mbox{Ei}(-x),
\quad
\int_{x}^{\infty}dy\frac{e^{-ny}}{y}=-\mbox{Ei}(-nx).
\label{eq79}
\end{equation}
\noindent
Finally, we obtain
\begin{equation}
F(ix)-F(-ix)=i\pi\,
\frac{(\ez-1)^2}{4\,(\ez+1)}\;x^2-i\gamma\, x^3+\mbox{O}(x^4),
\label{eq80}
\end{equation}
\noindent
where the unknown third order expansion coefficient is designated as $\gamma$.

Substituting Eq.~(\ref{eq80}) in Eq.~(\ref{eq19}) and using
Eq.~(\ref{eq16}), we find the free energy in the system
metal-dielectric in the form
\begin{equation}
{\cal F}(a,T)=E(a)-\frac{\hbar c}{32\pi^2a^3}
\left[\frac{\zeta(3)}{16\pi^2}\,
\frac{({\ee}_0-1)^2}{{\ee}_0+1}\;\tau^3-K_4\,\tau^4+
\mbox{O}(\tau^5)\right],
\label{eq81}
\end{equation}
\noindent
where $K_4\equiv\gamma/240$.

The Casimir pressure in the configuration of metal and dielectric
plates obtained from Eq.~(\ref{eq81}) is equal to
\begin{equation}
P(a,T)=
P_0(a)-\frac{\hbar c}{32\pi^2a^4}\Big[
K_4\,\tau^4+\mbox{O}(\tau^5)\Big].
\label{eq82}
\end{equation}
\noindent
The  direct application of  the Lifshitz
formula gives the expression for the pressure analogical to Eq.~(\ref{eq34}), 
\begin{eqnarray}
\hspace{-.3cm}
P(a,T)=-\frac{\hbar c\tau}{32\pi^2a^4}
\sum\limits_{l=0}^{\infty}
\left(1-\frac{\delta_{l0}}{2}\right)\!
\int_{\zeta_l}^{\infty}\!\!y^2dy
\left[\frac{r_{\|}(\zeta_l,y)}{e^{y}-r_{\|}(\zeta_l,y)}+
\frac{r_{\bot}(\zeta_l,y)}{e^{y}-r_{\bot}(\zeta_l,y)}\right]\!.
\label{eq83} 
\end{eqnarray}
\noindent
Using the Abel-Plana formula (\ref{eq15}), Eq.~(\ref{eq83}) can be 
represented in the form of Eqs.~(\ref{eq35}), (\ref{eq36})  
where
\begin{equation}
\Phi_{\|,\bot}(x)=\int_{x}^{\infty}dy
\frac{y^2\,r_{\|,\bot}(x,y)}{e^{y}-r_{\|,\bot}(x,y)}\, .
\label{eq84}
\end{equation}

Again, we deal first with $\Phi_{\bot}(x)$.
By adding and subtracting the asymptotic behavior of the integrated
function at small $x$,
\begin{equation}
\frac{y^2r_{\bot}(x,y)}{e^{y}-r_{\bot}(x,y)}=
\frac{1}{4}(\ez-1)x^2e^{-y}+\mbox{O}(x^3)\,,
\label{eq85}
\end{equation}
\noindent
and introducing the new variable $v=y/x$, the function $\Phi_{\bot}(x)$
can be identically rearranged and expanded in powers of $x$ as follows:
\begin{eqnarray}
&&\Phi_{\bot}(x)=\frac{1}{4}(\ez-1)x^2e^{-x}+x^3
\int_{1}^{\infty}dv\bigg[v^2
\sum\limits_{n=1}^{\infty}r_{\bot}^n(v)e^{-nvx}-
\frac{1}{4}(\ez-1)e^{-vx}\bigg]
\nonumber\\
&&\phantom{aaaaa}=\frac{1}{4}(\ez-1)x^2(1-x)+x^3
\int_{1}^{\infty}dv\left[
\frac{v^2r_{\bot}(v)}{1-r_{\bot}(v)}-
\frac{\ez-1}{4}\right]+\mbox{O}(x^4).
\label{eq87}
\end{eqnarray}
\noindent
The integral on the right-hand side of Eq.~(\ref{eq87})
is converging and can be simply calculated with the result
\begin{equation}
\Phi_{\bot}(x)=\frac{\ez-1}{4}x^2-\frac{1}{6}
\left(\ez\sqrt{\ez}-1\right)x^3+\mbox{O}(x^4).
\label{eq88}
\end{equation}

To deal with $\Phi_{\|}(x)$ we add and subtract in Eq.~(\ref{eq84})
the two first expansion terms of the integrated function in
powers of $x$,
\begin{eqnarray}
&&\Phi_{\|}(x)=
\int_{x}^{\infty}y^2dy\bigg[
\frac{r_0}{e^y-r_0}-
\frac{{\ee}_0r_0e^{-y}x^2}{y^2(\ez+1)\left(1-r_0e^{-y}\right)^2}
\bigg]
\label{eq89} \\
&&\phantom{aaaa}
+\int_{x}^{\infty}y^2dy
\bigg[\frac{r_{\|}(x,y)}{e^y-r_{\|}(x,y)}-
\frac{r_0}{e^y-r_0}+
\frac{{\ee}_0r_0e^{-y}x^2}{y^2(\ez+1)\left(1-r_0e^{-y}\right)^2}
\bigg].
\nonumber
\end{eqnarray}
\noindent
The asymptotic expansion of the first integral on the right-hand side
of Eq.~(\ref{eq89}) is given by
\begin{equation}
2\mbox{Li}_3(\rz)-\frac{\ez(\ez-1)}{2(\ez+1)}x^2+
\frac{1}{12}(\ez-1)(3\ez-2)x^3+\mbox{O}(x^4),
\label{eq90}
\end{equation}
\noindent
and of the second one by
\begin{equation}
\left[-\frac{1}{4}\ez(\ez-1)-\frac{1}{6}\ez(\ez\sqrt{\ez}-1)+
\frac{1}{2}\ez(\ez-1)\sqrt{\ez}\right]x^3+\mbox{O}(x^4).
\label{eq91}
\end{equation}
\noindent
By summing Eqs.~(\ref{eq90}) and (\ref{eq91}) we find 
\begin{eqnarray}
&&\Phi_{\|}(x)=
2\mbox{Li}_3(\rz)-\frac{\ez(\ez-1)}{2(\ez+1)}\,x^2
\label{eq92} \\
&&\phantom{aaaaa}
-\frac{1}{6}\big[(\ez-1)+(\ez\sqrt{\ez}-1)
-3\ez(\ez-1)\sqrt{\ez}\big]\,x^3+\mbox{O}(x^4).
\nonumber
\end{eqnarray}
\noindent
Finally we add Eq.~(\ref{eq92}) to Eq.~(\ref{eq88}) and arrive at
\begin{equation}
\Phi(ix)-\Phi(-ix) = -i\frac{2}{3}
\left(1-2\ez\sqrt{\ez}+\varepsilon_0^2\sqrt{\ez}\right)x^3
+\mbox{O}(x^4).
\label{eq93}
\end{equation}
\noindent
Substituting this in Eq.~(\ref{eq36}) and using Eq.~(\ref{eq35}),
the asymptotic expression for the Casimir pressure is obtained
\begin{equation}
P(a,T)=P_0(a)-\frac{\hbar c}{32\pi^2a^4}\left[
\frac{1}{360}\left(1-2\ez\sqrt{\ez}+\varepsilon_0^2\sqrt{\ez}\right)\,\tau^4
+\mbox{O}(\tau^5)\right].
\label{eq94}
\end{equation}
\noindent
Comparing Eqs.~(\ref{eq94}) and (\ref{eq82}), the explicit expression for
the coefficient $K_4$ reads
\begin{equation}
K_4=\frac{1}{360}\left(1-2\ez\sqrt{\ez}+\varepsilon_0^2\sqrt{\ez}\right).
\label{eq95}
\end{equation}

Now we are in a position to find the asymptotic behavior of the entropy in
the limit of low temperatures in the configuration of two parallel
plates one of which is metallic and the other one dielectric.
By calculating the negative derivative of Eq.~(\ref{eq81}) with
respect to temperature, one arrives at
\begin{eqnarray}
&&S(a,T)=
\frac{3k_B\zeta(3)({\ee}_0-1)^2}{128\pi^3a^2({\ee}_0+1)}\;\tau^2
\label{eq96} \\
& &\phantom{aaaaa}\times
\bigg[1-
\frac{8\pi^2({\ee}_0+1)\left(1-2\ez\sqrt{\ez}+
\varepsilon_0^2\sqrt{\ez}\right)}{135\zeta(3)(\ez-1)^2}
\;\tau+\mbox{O}(\tau^2)\bigg].
\nonumber
\end{eqnarray} 

This equation is in analogy to Eq.~(\ref{eq56}) obtained for the case
of two dielectrics. As is seen from Eq.~(\ref{eq96}), the entropy of the
Casimir and van der Waals interactions between metal and dielectric 
plates vanishes when the temperature goes to zero, i.e., the Nernst
heat theorem is satisfied. Note that the first term of order
$\tau^2$ on the right-hand side of Eq.~(\ref{eq96}) was already obtained in 
Ref.~\refcite{49}.
It is notable also that at low temperatures the entropy goes to zero
remaining positive. At the same time, as was shown in Ref.~\refcite{49},
at larger temperatures entropy is nonmonotonous and may take negative
values. This interesting property distinguishes the configuration
of metal and dielectric plates from two dielectric plates. In the latter
configuration the negative entropy appears only for nonphysical
dielectrics with anomalously large and frequency independent dielectric
permittivities.\cite{44}

Now let us consider the case $\tau\gg 1$,
i.e., high temperatures (large separations). In the same way as
for two dielectric plates, here the main contribution to the free energy
is given by the term with $l=0$ in Eq.~(\ref{eq75}),
\begin{equation}
{\cal F}(a,T)=\frac{\hbar c\tau}{64\pi^2a^3}
\int_{0}^{\infty}ydy\ln\left(1-r_0e^{-y}\right).
\label{eq96a}
\end{equation}
\noindent
Performing the
integration in Eq.~(\ref{eq96a}), we obtain
\begin{equation}
{\cal F}(a,T)=-\frac{k_B T}{16\pi a^2}
\mbox{Li}_3\left(r_0\right).
\label{eq96b}
\end{equation}

{}From this equation, for the Casimir pressure and entropy at
$\tau\gg 1$ it follows
\begin{equation}
P(a,T)=-\frac{k_B T}{8\pi a^3}
\mbox{Li}_3\left(r_0\right),
\qquad
S(a,T)=\frac{k_B}{16\pi a^2}
\mbox{Li}_3\left(r_0\right).
\label{eq96c}
\end{equation}
\noindent
The results (\ref{eq96b}) and (\ref{eq96c}) are analogous to
 (\ref{eq59b}) and (\ref{eq59c}) for two dielectric plates.

\section{THE PROBLEM ORIGINATING FROM THE ACCOUNT OF DIELECTRIC
DC CONDUCTIVITY}

In the previous section it was supposed that the static dielectric 
permittivity $\ez$ of the dielectric plate is finite. Now we will deal
with the configuration of metal and dielectric plates with included dc 
conductivity of the dielectric material. In doing so the permittivity of
the dielectric plate is given by
\begin{equation}
\tilde{\ee}(i\xi_l)=\ez+\frac{\beta(T)}{l},
\label{eq97}
\end{equation}
\noindent
where all notations were introduced in and below Eq.~(\ref{eq60}).
Thus, the reflection coefficients at zero frequency satisfy Eq.~(\ref{eq4})
for a plate made of ideal metal and Eq.~(\ref{eq61}) for a plate made of 
dielectric with included dc conductivity.
Let us find the low-temperature behavior of the Casimir entropy and
verify the consistency of the Lifshitz theory with thermodynamics in
this nonstandard situation.

For this purpose we substitute the dielectric permittivity (\ref{eq97})
in Eq.~(\ref{eq75}) instead of $\ez$ and find the Casimir energy
$\tilde{\cal F}(a,T)$ with included dc conductivity of a dielectric
plate. In the same way as in Sec.~3, it is convenient to separate
the zero-frequency term of $\tilde{\cal F}(a,T)$ and subtract and add 
the usual zero-frequency contribution for metal-dielectric plates
computed with the dielectric permittivity $\ez$,
\begin{eqnarray}
\tilde{\cal F}(a,T)
&=&\frac{k_BT}{16\pi a^2}
\int_{0}^{\infty}ydy\left[
\ln\left(1-e^{-y}\right)
-\ln\left(1-\rz e^{-y}\right)\right]
\nonumber\\
&
+&\frac{k_BT}{16\pi a^2}
\int_{0}^{\infty}ydy
\ln\left(1-\rz e^{-y}\right)
\label{eq98}\\
&
+&\frac{k_BT}{8\pi a^2}\sum\limits_{l=1}^{\infty}
\int_{\zeta_l}^{\infty}ydy\left\{
\ln\left[1-\tilde{r}_{\|}(\zeta_l,y)
e^{-y}\right]+
\ln\left[1-\tilde{r}_{\bot}(\zeta_l,y)
e^{-y}\right]
\right\}.
\nonumber
\end{eqnarray}
\noindent
Here the reflection coefficients $\tilde{r}_{\|,\bot}$ are
calculated with the permittivity (\ref{eq97}).
We expand the  third integral on the right-hand side
of Eq.~(\ref{eq98}) in powers of the small parameter
$\beta/l$. The zero-order contribution in
this expansion together with the second integral 
of Eq.~(\ref{eq98}) form the Casimir free energy
${\cal F}(a,T)$ calculated with  dielectric permittivity
$\ez$. The first integral on the 
right-hand side of Eq.~(\ref{eq98}) is calculated explicitly. 
As a result, Eq.~(\ref{eq98}) is rearranged to
\begin{equation}
\tilde{\cal F}(a,T)= {\cal F}(a,T)
-\frac{k_BT}{16\pi a^2}\left[\zeta(3)-
\mbox{Li}_3\left(\rz\right)\right]+Q(a,T),
\label{eq99}
\end{equation}
\noindent
where $Q(a,T)$ contains the first and higher-order contributions in the
expansion of the third integral on the right-hand side
of Eq.~(\ref{eq98}) in powers of $\beta/l$.
The explicit form  of the main first-order term in $Q(a,T)$ is
the following:
\begin{eqnarray}
\hspace{-.1cm}
&&Q_1(a,T)=\frac{k_BT}{8\pi a^2}
\sum\limits_{l=1}^{\infty}
\frac{\beta}{l}\int_{\zeta_l}^{\infty}
\frac{dy\,y^2e^{-y}}{\sqrt{y^2+\zeta_l^2({\ee}_0-1)}}
\left\{\frac{(2-{\ee}_0)\zeta_l^2-
2y^2}{\big[\sqrt{y^2+\zeta_l^2({\ee}_0-1)}+{\ee}_0y\,\big]^2}\,\,
\right.
\nonumber \\
&&\phantom{aaa}\left.
\times\frac{1}{1-r_{\|}(\zeta_l,y)e^{-y}}
-\frac{\zeta_l^2}{\big[\sqrt{y^2+\zeta_l^2({\ee}_0-1)}+y\,\big]^2}\,\,
\frac{1}{1-r_{\bot}(\zeta_l,y)e^{-y}}
\right\}.
\label{eq100}
\end{eqnarray}

In the same way as in Sec.~3, we
 expand the integrated function in Eq.~(\ref{eq100}) in powers of
$\tau$ (bearing in mind that $\zeta_l=\tau l$) and preserve
only the main contribution at
$\tau =0$:
\begin{eqnarray}
Q_1(a,T)&=&
-\frac{k_BT}{4\pi a^2}
\sum\limits_{l=1}^{\infty}
\frac{\beta}{l}\int_{\zeta_l}^{\infty}
\frac{dy\,y\rz e^{-y}}{(\varepsilon_0^2-1)\left(1-\rz e^{-y}\right)}
\label{eq101} \\
&=&
-\frac{k_BT\beta}{4\pi a^2(\ee_0^2-1)}
\sum\limits_{n=1}^{\infty}
\frac{r_0^{n}}{n^2}
\left[\sum\limits_{l=1}^{\infty}\frac{e^{-n\tau l}}{l}
+n\tau\sum\limits_{l=1}^{\infty}e^{-n\tau l}\right].
\nonumber 
\end{eqnarray}
\noindent
Dealing with this expression in the same way as with Eq.~(\ref{eq71}),
we arrive at
\begin{equation}
Q_1(a,T)\sim e^{-b/T}\ln T.
\label{eq102}
\end{equation}

The Casimir entropy in the configuration of metal and dielectric
plates with included dc conductivity of the dielectric plate is obtained
as minus derivative of Eq.~(\ref{eq99}) with respect to temperature, 
\begin{equation}
\tilde{S}(a,T)=S(a,T)+\frac{k_B}{16\pi a^2}\left[\zeta(3)-
\mbox{Li}_3\left(\rz\right)\right]-
\frac{\partial Q(a,T)}{\partial T}.
\label{eq103}
\end{equation}
\noindent
Using Eq.~(\ref{eq102}), the calculation of the limiting value at
 $T\to 0$ is straightforward:
\begin{equation}
\tilde{S}(a,0)=\frac{k_B}{16\pi a^2}\left[\zeta(3)-
\mbox{Li}_3\left(\rz\right)\right]>0.
\label{eq104}
\end{equation}
\noindent
{}From this equation it follows that the inclusion of the dc conductivity
of dielectric plate in the configuration metal-dielectric results in
a violation of the Nernst heat theorem. In the above this result was 
obtained for a metallic plate made of ideal metal. It can be shown that it
remains valid for a metallic plate made of real metal
with finite conductivity.

Thus, both configurations (two dielectric plates or one metal plate and
one dielectric) lead to the same conclusion that when the dc conductivity
is included in the model of dielectric response of the dielectric plate, 
the Lifshitz theory loses its consistency with thermodynamics. This confirms
the conclusion made in Sec.~3 that the actual properties of 
dielectric materials
at very low frequencies are in fact not related to the van der Waals
and Casimir forces.

\section{QUALITATIVE DISCUSSION OF
CASIMIR INTERACTION BETWEEN METAL AND SEMICONDUCTOR}

As was mentioned in the Introduction, semiconductors possess a wide
variety of electric and optical properties ranging from metallic to
dielectric. This opens the possibility to modulate the van der Waals 
and Casimir forces by changing the charge carrier density. Bearing in
mind the discussed above problems on the consistency of the Lifshitz
theory with thermodynamics, semiconductors can provide us with a test for 
the validity of different approaches. Thus, if for good dielectric the
dc conductivity does not play any real role in the van der Waals and
Casimir forces, the question arises on how much it should be increased
in order to become a relevant factor in the description of dispersion
forces.

In Ref.~\refcite{66} 
the Casimir force acting between two Si plates was
calculated using the simple analytic expression for Si dielectric 
permittivity as a function of frequency. The complete tabulated optical
data of Si were used in Ref.~\refcite{62} to calculate the van der Waals
interaction of different atoms with a Si wall. The first attempt to
measure the van der Waals force between a glass lens and a Si plate
and to modify it by light due to the change of carrier density was
undertaken in Ref.~\refcite{69}. 
However, glass is a dielectric and therefore
the electric forces due to localized point charges could not be
controlled. This might explain that no force change occured on
illumination at small separations where the effect should be most
pronounced.

The first measurements of the Casimir force between a gold coated sphere
and a single crystal Si plate were performed in Refs.~\refcite{50,51} by means
of the atomic force microscope. The experiments used a $p$-type B doped
Si plate of resistivity $\rho=0.0035\;\Omega\,$cm. The chosen resistivity 
of the plate is in some sense intermediate between the resistivity of 
metals (which is usually two or three orders of magnitude lower) and
the resistivity of dielectrics (it can be by about a factor of $10^5$ 
larger; for instance, high-resistivity ``dielectric'' Si has
the resistivity $\rho_0=1000\;\Omega\,$cm). Thus, the used Si plate had 
a relatively large absorption typical for semiconductors but it was also
enough conductive to avoid the accumulation of charges.\cite{51}
\begin{figure}[t]
\vspace*{-1.2cm}
\centerline{\hspace*{-4cm}\psfig{file=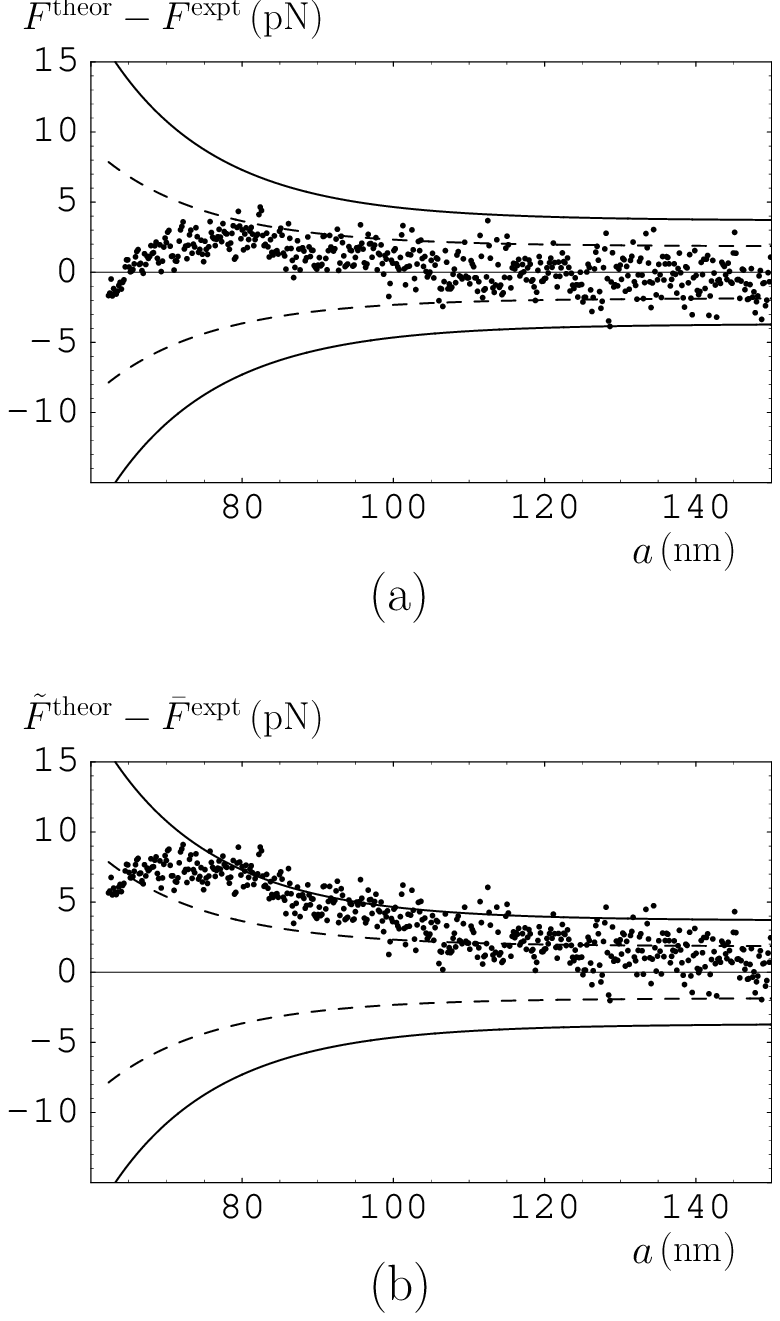,width=20cm}}
\vspace*{-13.6cm}
\caption{\label{fig2}
Differences of the theoretical and mean experimental Casimir
forces versus separation. Theoretical forces are computed
(a) for the Si plate used in experiment and (b) for dielectric Si.
Solid and dashed lines indicate 95 and 70\% confidence intervals,
respectively.}
\end{figure}

In Fig.~2, taken from Ref.~\refcite{51}, the differences of the theoretical 
and mean experimental Casimir forces acting between Au sphere and Si plate
are presented as functions of separation. In Fig.~2a the theoretical
force $F^{\rm theor}$ is computed using the Lifshitz formula and
the dielectric permittivity of a Si plate with the relatively low
resistivity $\rho$ used in experiment. This dielectric permittivity goes to
infinity as $\xi^{-1}$ with decreasing frequency (see the solid line in
Fig.~3). In Fig.~2b the theoretical Casimir force ${\tilde{F}}^{\rm theor}$
is computed using the dielectric permittivity of the Si plate made of
``dielectric'' Si with high resistivity $\rho_0$. The dielectric
permittivity of high-resistivity Si is shown by the dashed line in Fig.~3.
It is characterized   by the finite static value $\ee^{Si}(0)=11.67$.
The solid and dashed lines in Fig.~2a,b indicate the 95\% and 70\%
confidence intervals, respectively. As is seen from Fig.~2, the theoretical
approach using the dielectric permittivity of high-resistivity
``dielectric'' Si is excluded by experiment within the separation range
from 60 to 110\,nm at 70\% confidence. At the same time, the theory using the
dielectric permittivity of Si with a low resistivity $\rho$ is consistent
with experiment.
\begin{figure}[t]
\vspace*{-6.5cm}
\centerline{\hspace*{-1cm}\psfig{file=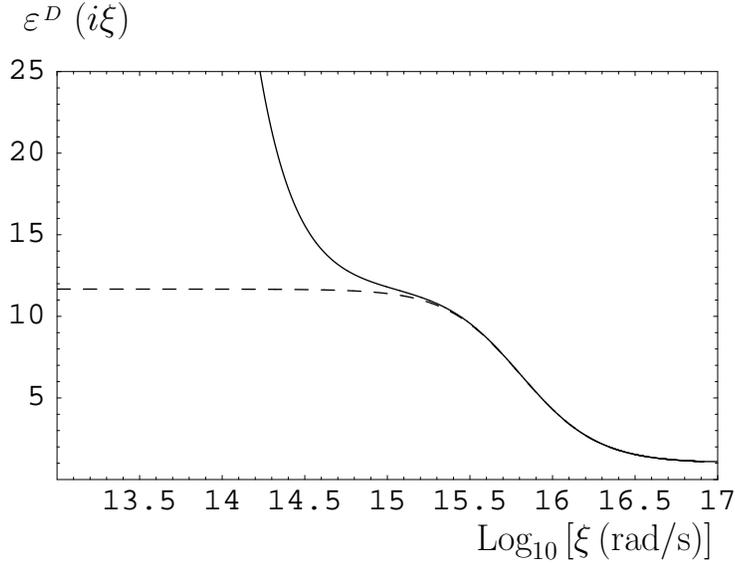,width=20cm}}
\vspace*{-13.6cm}
\caption{\label{fig3}
Dielectric permittivity of Si plate used in experiment along the
imaginary frequency axis (solid line). Dashed line shows the
dielectric permittivity of dielectric Si.}
\end{figure}

The above results suggest an approach on how to correctly determine
the possible role of the low-frequency conductivity properties in
dispersion forces. As is seen from Fig.~3, the dielectric permittivity of
low-resistivity Si (solid line) significantly departs from the dielectric
permittivity of ``dielectric'' Si in the region around the important
dimensional parameter of the problem, the characteristic 
frequency $c/(2a)\sim 10^{14}-10^{15}\,$rad/s.
 That is why the low-resistivity sample cannot be described at low frequencies
by the static dielectric permittivity of Si equal to
$\ee^{Si}(0)=11.67$. 
To describe it, the term $\beta(T)/l$, like in Eq.~(\ref{eq97}), should 
be added to $\varepsilon^{Si}(0)$. Note that in this case 
$\beta(T)/l>1$ at the first Matsubara frequencies with $l=1,\,2,\,3,\ldots$
and, thus, this quantity cannot be considered as a small parameter. 
At the same time, for a high-resistivity sample the
inclusion of the dc conductivity would lead to deviations from the
dashed line in Fig.~3 only at frequencies $\xi<10^8\,$rad/s, which are much
less than the characteristic frequency. This comparison permits to make
a conclusion in what experimental situations the conductivity properties 
of semiconductors at low frequencies should be taken into account and
when they should be omitted as being not related to dispersion forces.

The future experiments on the modification of semiconductor charge
carrier density by laser light\cite{Umar06} 
will bring a more clear understanding
of this problem on the connection between the low-frequency material
properties and dispersion forces.

\section{DOES SPATIAL DISPERSION LEAD TO AN IMPORTANT IMPACT ON
THERMAL CASIMIR FORCE?}

The presented above new analytic results on the low-temperature
behavior of the Casimir free energy, pressure and entropy between
dielectrics or between dielectric and metal are based on the
conventional Lifshitz theory which describes dielectric materials
by means of the frequency dependent dielectric permittivity.
In fact, the assumption that the material of the plates possesses only
the frequency dispersion means that the components of electric
displacement are connected with the components of electric field by
the relation
\begin{equation}
\mbox{{{$D$}}}_{k}(\mbox{\boldmath$r$},\omega)=
\varepsilon_{kl}(\mbox{\boldmath$r$},\omega)
\mbox{{$E$}}_{l}(\mbox{{\boldmath{$r$}}},\omega).
\label{eq105}
\end{equation}
\noindent
This equation is central in all different derivations of the Lifshitz
formula (see, e.g., 
Refs.~\refcite{2,6},\ \refcite{29}--\refcite{30},\ \refcite{57,57a,70}). 
The effects of spatial 
nonlocality (spatial dispersion) are in fact essential only at shortest 
separations between the plates comparable with atomic dimensions and
also for metals
at sufficiently large separations (typically of about 2--3$\,\mu$m)
in the frequency region of the anomalous skin effect. The Casimir force
in the latter region was described by the Lifshitz theory reformulated 
in terms of the Leontovich impedance.\cite{71}

Recently the spatial dispersion came to the attention in connection
with the thermal Casimir force.\cite{52,53,54,55,56} 
In particular, it was claimed\cite{52,53} 
that for real metals at any separation the account of spatial 
dispersion leads to practically the same result (\ref{eq6}) for the 
reflection coefficients at zero frequency as was obtained earlier using 
the Drude model dielectric function (\ref{eq5}). This conclusion, if it
is correct, not only returns us to the contradiction with 
experiment\cite{26,28,28a} 
but also casts doubts on all results obtained by means of the
conventional Lifshitz theory accounting for only the frequency
dispersion. It is natural when the spatial dispersion contributes a
small fraction of a percent as it is generally believed in numerous
applications of the Lifshitz theory. It is, however, quite another
matter when the account of the spatial dispersion results in some
``dramatic effects'',\cite{52} i.e., in several hundred times larger thermal
correction than is obtained in the local case. Below we demonstrate that the
conclusions of Refs.~\refcite{52} -- \refcite{56} 
are in fact not reliable because they use the
Lifshitz theory of dispersion forces outside of its application
range.\cite{72}

To find the electromagnetic modes associated with an empty gap
between the plates, Refs.~\refcite{52} -- \refcite{56}
use the standard continuity boundary conditions,
\begin{equation}
E_{1t}=E_{2t}, \quad
B_{1n}=B_{2n}, \quad
D_{1n}=D_{2n},\quad
B_{1t}=B_{2t}, 
\label{eq106}
\end{equation}
\noindent
which are commonly applied in the derivation of the Lifshitz formula
for spatially local nonmagnetic materials.  
Here {\boldmath$B$} is the magnetic induction, 
{\boldmath$n$} is the normal to the boundary directed
inside the medium, the subscripts $n,t$ refer to the normal
and tangential components, respectively,
the subscript 1 refers to the vacuum and subscript
2 to the plate material. 
In Refs.~\refcite{52} -- \refcite{56} 
the spatial dispersion is described by the 
longitudinal and transverse dielectric permittivities depending
on the wave vector and frequency: 
$\ee_{kl}=\ee_{kl}({\mbox{\boldmath$q$}},\omega)$. However, 
as is shown below, in the theory of
the Casimir effect both the boundary conditions (\ref{eq106}) and 
permittivities 
$\ee_{kl}=\ee_{kl}({\mbox{\boldmath$q$}},\omega)$ are inapplicable.

We start from the boundary conditions and recall the set of
Maxwell equations in a metal describing the Casimir effect,
\begin{equation}
\mbox{rot\boldmath$E$}+
\frac{1}{c}\frac{\partial{\mbox{\boldmath$B$}}}{\partial t}=0,
\quad
\mbox{div}\mbox{\boldmath$D$}=0,
\quad
\mbox{rot\boldmath$B$}-
\frac{1}{c}\frac{\partial{\mbox{\boldmath$D$}}}{\partial t}=0
\quad
\mbox{div}\mbox{\boldmath$B$}=0.
\label{eq107}
\end{equation}
\noindent
Equations (\ref{eq107}) do not contain any external, i.e., independent
on {\boldmath${E}$}, {\boldmath${D}$} and 
{\boldmath${B}$}, current or charge densities.
The definition of the electric displacement is
\begin{equation}
\frac{\partial{\mbox{{\boldmath$D$}}}}{\partial t}=
\frac{\partial\mbox{{\boldmath$E$}}{}}{\partial t}+
4\pi{\mbox{{\boldmath$i$}}},
\label{eq108}
\end{equation}
\noindent
where the volume current {\boldmath${i}$} is induced by
{\boldmath${E}$} and {\boldmath${B}$} and takes 
into account the conduction electrons. 

In electrodynamics with spatial dispersion the electric field and magnetic 
induction are finite at the boundary surfaces, whereas the electric
displacement can tend to infinity.\cite{73} 
Then, integrating Eqs.~(\ref{eq107})
over the thickness of the boundary layer as is done in Ref.~\refcite{74},
we reproduce the first two conditions in Eq.~(\ref{eq106}) and arrive
at the modified third and fourth conditions,\cite{73,75}
\begin{equation}
E_{1t}=E_{2t}, \quad
B_{1n}=B_{2n}, 
\quad
D_{2n}-D_{1n}=4\pi\sigma,\quad
[\mbox{\boldmath{$n$}}\times({\mbox{\boldmath{$B$}}}_2-
{\mbox{\boldmath{$B$}}}_1)]=\frac{4\pi}{c}\mbox{\boldmath{$j$}},
\label{eq109}
\end{equation}
\noindent
where the induced surface charge and current densities are
given by
\begin{equation}
\sigma=\frac{1}{4\pi}\int_{1}^{2}
\mbox{div}[\mbox{\boldmath{$n$}}\times
[\mbox{\boldmath{$D$}}\times\mbox{\boldmath{$n$}}]]dl,
\quad
\mbox{\boldmath{$j$}}=\frac{1}{4\pi}\int_{1}^{2}
\frac{\partial \mbox{\boldmath{$D$}}}{\partial t}dl.
\label{eq110}
\end{equation}
\noindent
Note that the boundary conditions (\ref{eq109}), (\ref{eq110})
are obtained from the macroscopic Maxwell equations for
physical fields. They should not be mixed 
with the boundary conditions arising in perturbative theories and for the
fictitious fields (see below).

In linear electrodynamics for a medium with time-independent
properties without spatial dispersion the material equation connecting
the electric displacement and electric field takes the form
\begin{equation}
\mbox{$D$}_{k}(\mbox{\boldmath$r$},t)=\int_{-\infty}^{t}
\hat{\varepsilon}_{kl}(\mbox{\boldmath$r$},t-t^{\prime})
\mbox{$E$}_{l}(\mbox{\boldmath$r$},t^{\prime})
dt^{\prime}.
\label{eq111}
\end{equation}
\noindent
 According to this equation, the electric 
displacement at a point
{\boldmath$r$} and moment $t$ is determined by the electric
field at the same point {\boldmath$r$} at different moments
$t^{\prime}\leq t$ (the spatial dispersion is absent but the temporal 
may be present). It is easily seen that the substitution of
Eq.~(\ref{eq111}) in Eq.~(\ref{eq110}) leads to $\sigma=0$,
 {\boldmath$j$}$=0$ and, as a result,
 the boundary conditions (\ref{eq109})
coincide with the standard continuity conditions (\ref{eq106}).
It is unjustified, however, to use  conditions
(\ref{eq106}) in the presence of spatial dispersion.
In Refs.~\refcite{73,76} a few examples are presented illustrating that
in this case  neither $\sigma$ nor {\boldmath$j$} is equal to zero.

We now turn to a discussion of the use of dielectric permittivity
$\ee_{kl}(\mbox{\boldmath$q$},\omega)$ in the theory of the Casimir
effect with account of spatial dispersion.
In the presence of only frequency dispersion, it is possible to perform the
Fourier transformation of the fields
\begin{equation}
\mbox{{\boldmath{$E$}}}({\mbox{\boldmath$r$}},t)
=\int_{-\infty}^{\infty}
\mbox{\boldmath{$E$}}(\mbox{{\boldmath{$r$}}},\omega)
e^{-i\omega t}d\omega,
\quad
\mbox{{\boldmath{$D$}}}({\mbox{\boldmath$r$}},t)
=\int_{-\infty}^{\infty}
\mbox{\boldmath{$D$}}(\mbox{{\boldmath{$r$}}},\omega)
e^{-i\omega t}d\omega,
\label{eq112}
\end{equation}
\noindent
in Eq.~(\ref{eq111}) and arrive at Eq.~(\ref{eq105}) where
\begin{equation}
\varepsilon_{kl}(\mbox{\boldmath$r$},\omega)=
\int_{0}^{\infty}
\hat{\varepsilon}_{kl}(\mbox{{\boldmath{$r$}}},\tau)
e^{i\omega\tau}d\tau
\label{eq113}
\end{equation}
\noindent
is the frequency-dependent dielectric permittivity and
$\tau\equiv t-t^{\prime}$.
In fact Eqs.~(\ref{eq105}), (\ref{eq112}) and (\ref{eq113}) are used
in parallel with the boundary conditions (\ref{eq106}) in all
derivations of the Lifshitz formula.\cite{2,6,29,29a,30,57,57a,70}

If the material of the plates is characterized not only by temporal
but also spatial dispersion, Eq.~(\ref{eq111}) is generalized to
\begin{equation}
\mbox{{{$D$}}}_{k}(\mbox{\boldmath$r$},t)=\int_{-\infty}^{t}
dt^{\prime}
\int d{\mbox{\boldmath$r$}}^{\prime}
\hat{\varepsilon}_{kl}(\mbox{{\boldmath{$r$}}},
{\mbox{\boldmath$r$}}^{\prime},t-t^{\prime})
\mbox{{{$E$}}}_{l}({\mbox{\boldmath{$r$}}}^{\prime},t^{\prime}).
\label{eq114}
\end{equation}
\noindent
If the material medium were uniform in space (i.e., all points were
equivalent), the kernel $\hat{\varepsilon}$ would not depend 
on {\boldmath$r$} and {\boldmath$r$}${}^{\prime}$ separately,
as  in Eq.~(\ref{eq114}), but on the difference 
{\boldmath$R\equiv r-r^{\prime}$}.
In this case, by performing the Fourier transformation,
\begin{eqnarray}
&&\mbox{{\boldmath{$E$}}}({\mbox{\boldmath$r$}},t)
=\int_{-\infty}^{\infty}d\omega\int d{\mbox{\boldmath$q$}}
\mbox{\boldmath{$E$}}(\mbox{{\boldmath{$q$}}},\omega)
e^{i({\mbox{\scriptsize\boldmath$qr$}}-\omega t)},
\nonumber \\
&&
\mbox{{\boldmath{$D$}}}({\mbox{\boldmath$r$}},t)
=\int_{-\infty}^{\infty}d\omega\int d{\mbox{\boldmath$q$}}
\mbox{\boldmath{$D$}}(\mbox{{\boldmath{$q$}}},\omega)
e^{i({\mbox{\scriptsize\boldmath$qr$}}-\omega t)},
\label{eq115}
\end{eqnarray}
\noindent
and substituting it in
Eq.~(\ref{eq114}), one could introduce the dielectric permittivities
\begin{equation}
\varepsilon_{kl}(\mbox{\boldmath$q$},\omega)=
\int_{0}^{\infty}d\tau\int d\mbox{\boldmath$R$}\,
\hat{\varepsilon}_{kl}(\mbox{{\boldmath{$R$}}},\tau)
e^{-i(\mbox{\scriptsize\boldmath$qR$}-\omega\tau)},
\label{eq116}
\end{equation}
\noindent
as Refs.~\refcite{52}--\refcite{56} do, 
and rearrange Eq.~(\ref{eq114}) to the form
\begin{equation}
\mbox{{{$D$}}}_{k}(\mbox{\boldmath$q$},\omega)=
\varepsilon_{kl}(\mbox{\boldmath$q$},\omega)
\mbox{{$E$}}_{l}(\mbox{{\boldmath{$q$}}},\omega).
\label{eq117}
\end{equation}

In the Casimir effect, however, the material medium is not uniform
due to the presence of a macroscopic gap between the two plates
(half spaces). Because of this, the assumption that the 
 kernel $\hat{\varepsilon}$ depends only on
{\boldmath$R$} and $\tau$ is wrong. As a result, it is not possible
to introduce the dielectric permittivity
$\varepsilon_{kl}(\mbox{\boldmath$q$},\omega)$ depending on the wave
vector and frequency. In fact, for systems with spatial dispersion 
in the presence of boundaries the kernel $\hat{\varepsilon}$ depends
not only on  {\boldmath$R$} and $\tau$ 
but also on the distance from the boundary.\cite{73} 
In this complicated situation the following
approximate phenomenological approach is sometimes applicable.\cite{73} 
For electromagnetic waves with
a wavelength $\lambda$ the kernel
$\hat{\varepsilon}(\mbox{\boldmath$r$},
\mbox{\boldmath$r$}^{\prime},\tau)$ in Eq.~(\ref{eq114})
differs essentially from zero only
in a certain vicinity of the point {\boldmath$r$} with
characteristic dimensions $l\ll\lambda$ (for nonmetallic condensed media
$l$ is of the order of the lattice constant). 
Then it is reasonable to assume that $\hat{\varepsilon}$ is a function of 
{\boldmath$R$}$=${\boldmath$r$}--{\boldmath$r$}${}^{\prime}$,
except for a layer of thickness $l$ adjacent to the
boundary surface. If one is mostly interested in bulk phenomena
and neglects the role and influence of a
subsurface layer, the quantity 
$\varepsilon_{kl}(\mbox{\boldmath$q$},\omega)$ may be employed
as a reasonable approximation.

This approximate phenomenological approach is widely applied in the
theoretical investigation of the anomalous skin effect 
(see, e.g., Ref.~\refcite{77}).
Note that in Ref.~\refcite{77} 
some kind of fictitious infinite system was introduced 
and electromagnetic fields in this system are discontinuous on the
boundary surface. This discontinuity should not be confused with the
discontinuity of physical fields of a real system in the presence of
spatial dispersion described by Eqs.~(\ref{eq109}) and (\ref{eq110}).
(There is also another approach to the description of the anomalous
skin effect in polycrystals using the generalizations of the local
Leontovich impedance which takes into account the shape of Fermi
surface\cite{78}). The frequency and wave vector dependent dielectric
permittivity in the presence of boundaries is also approximately applied
in the theory of radiative heat transfer\cite{79} or in the study of
electromagnetic interaction of molecules with metal surfaces.\cite{80}
In all these applications the boundary effects are usually taken into
account by the boundary conditions (\ref{eq109}) supplemented by so called
``additional boundary conditions''.

It is unlikely, however, that the approximate phenomenological approach using
such quantity as $\varepsilon_{kl}(\mbox{\boldmath$q$},\omega)$ in the
presence of boundaries would be applicable in the theory of the Casimir force
where the boundary effects on the zero-point electromagnetic oscillations
are of prime importance. It is notable also that this approach faces serious 
theoretical difficulties including the violation of the law of conservation
of energy.\cite{81} It is then not surprising that the application
of this approach in Refs.~\refcite{52,53} results in the contribution to the
Casimir free energy from the transverse electric mode which is in
contradiction with experiment.\cite{26,28,28a}

One more shortcoming of Refs.~\refcite{52}--\refcite{56} 
is that they substitute the
dielectric permittivity $\varepsilon_{kl}(\mbox{\boldmath$q$},\omega)$,
depending on both wave vector and frequency, into the conventional
Lifshitz formula derived in the presence of only temporal dispersion. 
In the famous review paper\cite{82} it has been noticed, however, that with the
inclusion of spatial dispersion the free energy of a fluctuating field
takes the form ${\cal F}={\cal F}_L+\Delta{\cal F}$, where ${\cal F}_L$
is given by the conventional Lifshitz expression derived in a spatially
local case and written in terms of the Fresnel reflection coefficients,
and $\Delta{\cal F}$ is an additional term which can be expressed in terms 
of the thermal Green's function of the electromagnetic field and
polarization operator. Review\cite{82} calls as not reliable the results of,
e.g., Ref.~\refcite{83} obtained by the substitution of dielectric permittivity
$\varepsilon_{kl}(\mbox{\boldmath$q$},\omega)$, taking account of spatial
dispersion, into the conventional Lifshitz formula. It can be true that
the Lifshitz formula written in terms of general reflection coefficients
is applicable in both spatially local and nonlocal situations.
However, as far as the exact reflection coefficients in a spatially
nonlocal case are unknown, the use of some approximate
phenomenological models, elaborated in literature
for applications different than the Casimir effect, may lead to incorrect
results for $\Delta{\cal F}$ and create inconsistencies with experiment.

To conclude this section,  the results of Refs.~\refcite{52}--\refcite{56} 
on the influence of
spatial nonlocality on the Casimir interaction are shown to be not reliable.
Although at present there is no fundamental theory of the thermal Casimir
force incorporating spatial dispersion, there is no reason to expect
that it can play any significant role in the frequency region of
infrared optics (experimental separations) or normal skin effect (i.e.,
at separations between plates greater than 4--5$\,\mu$m).

\section{CONCLUSIONS AND DISCUSSION}

In the foregoing we have presented the derivation of 
analytic asymptotic expressions for
the free energy, pressure and entropy of the Casimir interaction between
two dielectric plates and between metal and dielectric plates at low
and high temperatures. It was shown that the low-temperature behavior of the
Casimir interaction between dielectrics and between dielectric and metal 
is determined by the static dielectric permittivities of nonpolar
dielectrics.
The obtained results were shown to be in agreement with thermodynamics
when the static dielectric permittivities of dielectrics are finite.
In particular, the entropy of the Casimir interaction goes to zero when 
the temperature vanishes, i.e., the Nernst heat theorem is satisfied.
This demonstrates the consistency of the original Lifshitz's
approach to the van der Waals forces between dielectrics which
disregards the small conductivity of dielectrics at constant current. 

The second important result shown above is that the inclusion of
the dc conductivity of dielectrics into the model of dielectric
response leads not to some small corrections to the characteristics
of the Casimir interaction, as one could expect, but makes the
Lifshitz theory inconsistent with thermodynamics leading to the 
violation of the Nernst heat theorem.  This reveals that real
material properties at very low, quasistatic frequencies are in fact
not related to the phenomenon of the van der Waals and Casimir forces
which is actually determined by  sufficiently high frequencies.
In this case the zero-frequency contribution to the Casimir force
should be understood not literally but as  analytic continuation to
zero of the material physical behavior in the region around
the characteristic frequency.

The presented results provide a basis for the calculation  of the
van der Waals and Casimir forces between real materials. Such
calculations are much needed for numerous applications of the Casimir
force discussed in the Introduction, in particular for the applications
in nanotechnology and for constraining predictions of fundamental
physical theories beyond the Standard Model. 
Bearing in mind that semiconductors are the main
constituent materials in nanotechnological devices, it is a subject
of high priority to understand the Casimir effect with semiconductor
boundaries. In this connection we have discussed new experimental results
and theoretical ideas on the Casimir interaction between a metal sphere
and semiconductor plate. It was stressed that by changing the charge
carrier density in the semiconductor it is possible to bring it in
different intermediate states between metallic and dielectric. In this 
case the problem arises when the conductivity
properties of semiconductor 
are not related to dispersion forces and when they are becoming
relevant. A criterion for the resolution of this problem was
formulated based on the relation between the typical frequency 
at which the dc conductivity properties come into play
and the characteristic frequency of the
Casimir effect. In fact the thermal Casimir interaction between
semiconductors remains an open question and much work should be done
both in experiment and theory to gain a better insight into this
subject.

The last major problem discussed in this review is whether or not 
the spatial dispersion influences essentially the thermal Casimir
force between real materials. In the present literature there is no
agreement on this subject. We adduced arguments in favor of the
statement that in the region of experimental separations the
influence of the spatial dispersion on the Casimir force is negligible 
small. The statements on the opposite, contained in the literature, were
shown to be not reliable because they are obtained by the application
of the Lifshitz theory outside of its application range.
At the same time it was ascertained that at the moment there is no
consistent fundamental theory of the van der Waals and Casimir forces taking
the spatial dispersion into account. This is the problem to solve
in the foreseeable future. 

\section*{ACKNOWLEDGMENTS}
G.L.K. and V.M.M. are grateful to the Center of Theoretical Studies and
the Institute for Theoretical
Physics, Leipzig University for their kind hospitality. 
This work was supported by Deutsche Forschungsgemeinschaft grant 
436\,RUS\,113/789/0-2.

\end{document}